\documentclass[12pt] {article}
\pdfoutput=1
\usepackage[hyphens,spaces,obeyspaces]{url}
\usepackage{setspace}

\usepackage{epsfig}
\usepackage{amssymb}
\usepackage{amsmath}
\usepackage{multirow}
\usepackage{setspace}
\usepackage{datetime}
\usepackage{amsmath}
\usepackage{algorithm}
\usepackage{algpseudocode}
\usepackage{graphicx}
\usepackage{lipsum}
\usepackage{acronym}
\usepackage{array}
\usepackage{graphicx}
\usepackage{cite}
\usepackage{graphicx}
\usepackage{epstopdf}
\usepackage{caption}
\usepackage{balance}
\usepackage{fancybox}
\usepackage{colortbl}
\usepackage{array}
\usepackage{mystyle}
\usepackage{multirow}
\usepackage[obeyspaces,hyphens,spaces]{url}
\usepackage{threeparttable}
\usepackage[english]{babel}
\usepackage{longtable}
\usepackage{listings, lstautogobble}
\usepackage{booktabs}
\usepackage{threeparttable}
\usepackage[table,xcdraw]{xcolor}
\usepackage{wrapfig}
\usepackage{subfigure}
\usepackage{paralist}
\usepackage{ifsym}

\usepackage{amsthm}
\AtBeginEnvironment{thm}{\footnotesize}
\AtBeginEnvironment{defn}{\footnotesize}
\providecommand{\keywords}[1]{\textbf{\textit{Keywords---}} #1}

\usepackage{stackengine}
\newsavebox\mybox

\usepackage{hyperref}
\usepackage[framemethod=tikz]{mdframed}
\usepackage{enumitem}
\theoremstyle{definition}
\AtBeginEnvironment{thm}{\footnotesize}
\AtBeginEnvironment{defn}{\footnotesize}

{\begin{list}{}%
         {\setlength{\leftmargin}{#1}}%
         \item[]%
}
{\end{list}}

\newdateformat{monthyeardate}{%
  \monthname[\THEMONTH], \THEYEAR}
\begin{document}
\sloppy

\title{\large{\textbf{Optical Stochastic Computing Architectures
Using Photonic Crystal Nanocavities } }}
\author{\normalsize
Hassnaa El-Derhalli$^{1}$, Lea Constans$^{2,3}$, S\'{e}bastien Le Beux$^{1}$,\\ \normalsize Alfredo De Rossi$^{2}$, Fabrice Raineri$^{3,4}$ and Sofi\`ene~Tahar$^{1}$\vspace*{1em}\\
\small $^{1}$Department of Electrical and Computer Engineering,\\
\small Concordia University, Montr\'eal, QC, Canada \\
\small\{h\_elderh,slebeux,tahar\}@ece.concordia.ca 
 \vspace*{0.5em}\\
\small$^{2}$Thales Research and Technology, Palaiseau, France\\
\small alfredo.derossi@thalesgroup.com
 \vspace*{0.5em}\\
\small $^{3}$Centre de Nanosciences et de Nanotechnologies, CNRS,\\
\small Université Paris-Saclay, Palaiseau, France\\
\small lea.constans@universite-paris-saclay.fr
 \vspace*{0.5em}\\
\small $^{4}$Centre de Nanosciences et de Nanotechnologies,\\
\small Université de Paris, Palaiseau, France\\
\small fabrice.raineri@c2n.upsaclay.fr 

 \vspace*{4em}\\
\textbf{\normalsize TECHNICAL REPORT}\\
\vspace{-20pt}
\date{\normalsize February 2021}
}
\maketitle

\newpage
\begin{abstract}
Stochastic computing allows a drastic reduction in hardware complexity using serial processing of bit streams. While the induced high computing latency can be overcome using integrated optics technology, the design of realistic optical stochastic computing architectures calls for energy efficient switching devices. Photonics Crystal (PhC) nanocavities are $\mu m^2$ scale devices offering 100fJ switching operation under picoseconds-scale switching speed. Fabrication process allows controlling the Quality factor of each nanocavity resonance, leading to opportunities to implement architectures involving cascaded gates and multi-wavelength signaling. In this report, we investigate the design of cascaded gates architecture using nanocavities in the context of stochastic computing. We propose a transmission model considering key nanocavity device parameters, such as Quality factors, resonance wavelength and switching efficiency. The model is calibrated with experimental measurements. We propose the design of XOR gate and multiplexer. We illustrate the use of the gates to design an edge detection filter. System-level exploration of laser power, bit-stream length and bit-error rate is carried out for the processing of gray-scale images. The results show that the proposed architecture leads to 8.5nJ/pixel energy consumption and 512ns/pixel processing time.

\end{abstract}

\newpage

\section{Introduction}
Stochastic computing trades off computing accuracy with energy consumption. The probabilistic presentation of the data and the serial processing of bit streams allow for reduced hardware complexity and high energy efficient design~\cite{alaghi2013survey}. Stochastic computing is suitable for error tolerant applications, such as image processing~\cite{li2011using}. It is also resilient to soft and transient errors since it does not involve weighted binary numbers~\cite{alaghi2017promise}, i.e., the weight of all bits in a stochastic bit stream is the same. However, the intrinsic high latency, induced by the serial computation, is the main limitation of this approach. On a technological side, integrated optics technology, which provides high speed signal propagation and high bandwidth~\cite{sun2015single}, has been widely used to accelerate computing architectures, such as optical neural networks~\cite{shen2017deep} and reconfigurable optical processors~\cite{anderson2020roc}. 

Silicon photonics devices, such as MZI and MRR have been widely investigated in the design of optical computing architectures~\cite{ribeiro2016demonstration, xu2011reconfigurable}. In these approaches, optical signals are modulated by electrical signals, which calls for costly electronics-to-optical and optical-to-electronics (EO/OE) converters for the design of large-scale architectures. To cope with this limitation, the design of all-optical gates using MRR has been investigated in~\cite{van2002optical}. The switching operation is obtained by applying a high power (typically few mW) optical \textit{control} signal in order to modulate a lower power optical \textit{data} signal (typically few 100s $\mu$W). In MRRs, this is achieved by injecting control and data signals on different resonant wavelength: the wavelength detuning obtained from the control signal will modify the transmission of the data signal. The difference in transmission between optical signals representing data ‘1’ and ‘0’ is called Extinction Ratio (ER). This way, the data signals remain in the optical domain during their processing from the inputs to the outputs, which prevent from the need for EO/OE converters. Therefore, all-optical architectures have the potential to operate at higher speeds compared to optical architectures involving electrically controlled devices. However, to trigger non-linear effects needed for the all-optical computing, one has to take into account the wavelength detuning achievable in the MRR, which mostly depends on the Quality factor (\textit{Q} factor). Since the \textit{Q} factor is intrinsically the same for all resonances, the modulation obtained on the data signal is necessary limited by the shift triggered by the control signal. Photonic Crystal (PhC) nanocavities do not share this limitation since each resonance can show a different \textit{Q} factor. Hence, using such a device can lead to extinction ratio unreachable with MRR, which is essential for the design of computing architectures involving cascaded gates. Furthermore, PhC demonstrates 10ps switching speed, 100fJ switching energy consumption and 10$\times$ compactness compared to MRRs~\cite{bazin2014ultrafast}, which make the devices an ideal candidate for all-optical computing architectures.

The design of all-optical gates is necessary to implement all-optical computing architectures. In the context of stochastic computing, the design of all-optical XOR gate and Multiplexer (MUX) is essential since they represent an absolute value subtractor and an adder, respectively. The implementation of an architecture that involves cascaded gates, such as stochastic edge detection with cascaded multiplexers, in optical domain is challenging. It requires a device with different \textit{Q} factors and wavelength detuning to transmit a group of signals propagating at multiple wavelengths. The design of such architecture involves a large design space to explore at both device and system levels, such as \textit{Q} factors, resonance wavelength, and wavelength detuning.

In this work, we investigate the use of PhC nanocavities to design all-optical cascaded gates for stochastic computing architectures. For this purpose, we develop all-optical XOR gate and multiplexer (MUX) using nanocavities. We propose a transmission model of the nanocavities taking into account \textit{Q} factors and resonance wavelengths, which allows to explore the design space. As a case study, we implement a Sobel edge detection filter, which involves cascaded XOR gate and MUX for absolute value subtraction and addition. The design of the cavities is explored to trade off power consumption, computing accuracy and processing time. System-level evaluation is carried out through the processing of images under various Bit Stream Lengths (\textit{BSL}) and laser powers.

The rest of the report is organized as follows: Section 2 presents an overview of image processing filters implemented using stochastic computing and introduces existing optical computing architectures. In Section 3, we introduce the design of NOT gate, XOR gate and MUX using PhC nanocavity, and present the proposed transmission model of the device. Section 4 illustrates the design of a stochastic edge detection filter-based architecture using Sobel operators. The analytical model used to estimate the required lasers power and evaluate the computing accuracy is introduced in Section 5. In Section 6, the simulation results are presented. Finally, we conclude the report and present future work.

\section{Background and Related Work}
\subsection{Stochastic Computing}

Computations in stochastic computing are performed on probabilities instead of weighted binary numbers. A Stochastic Number Generator (SNG) generates bit streams, where the ratio of the number of 1's to the \textit{BSL} indicates the probability~\cite{alaghi2013survey}. Therefore, the result is approximated, and the accuracy is enhanced by increasing the \textit{BSL}. Stochastic computing is characterized by reduced hardware complexity. For example, an addition can be implemented using a 2-1 MUX.

Different architectures were proposed to perform stochastic computations. The reconfigurable architecture in~\cite{qian2010architecture}, can execute any arbitrary single input function. It relies on transforming the targeted function to its equivalent Bernstein polynomial function. In~\cite{yuan2016high}, the architecture is designed to implement high accuracy FIR filters by proposing non-scaled stochastic adder. The design of Low-Density Parity Check (LDPC) decoding~\cite{lee20147} in communication domain can be implemented using stochastic circuits to perform parity checking and equality checking~\cite{alaghi2017promise}. 

In the context of neural network, a deep neural network (DNN) relying on the approximation of any real number using an integer stochastic stream is proposed in~\cite{ardakani2017vlsi}. It results in 45\% and 62\% reduction in area and latency, respectively, compared to the state-of-the-art stochastic architecture. A convolutional neural network (CNN) relying on hybrid bit stream-binary is proposed in~\cite{faraji2019energy}. The design of the first layer is based on low-discrepancy deterministic bit streams for accurate and fast computing. The results show 19× area reduction and 16$\times$ power saving compared to the non-pipelined fixed point binary design.
 
In~\cite{alaghi2013stochastic}, the design of stochastic edge detection filter is proposed. It is based on Robert's cross operator, shown in Figure~\ref{fig:electrica_edge_detection}, where two 2$\times$2 filters are applied to an image in order to find the gradient vector at each pixel. The filters rely on absolute value subtraction and addition that are implemented using XOR gate and multiplexer (MUX), respectively, as detailed in the following:

\begin{figure}[!t]
\centering
\captionsetup{justification=centering}
\includegraphics[width=0.8\textwidth]{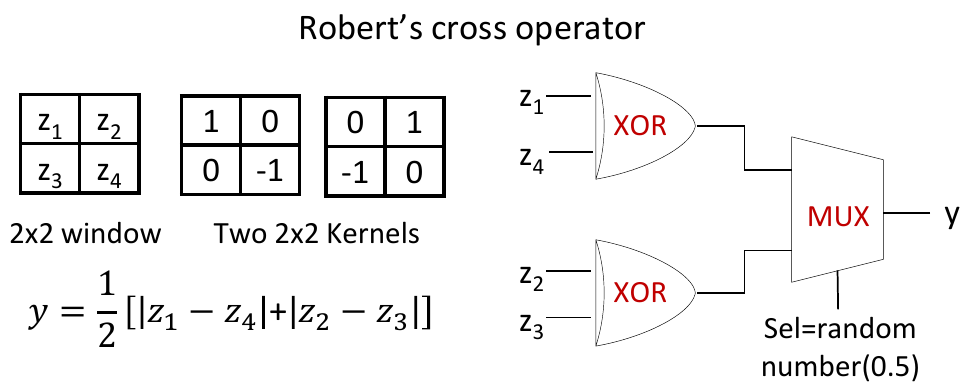}
\caption[Stochastic implementation of edge detection filter using Robert’s cross operator ]{Stochastic implementation of edge detection filter using Robert’s cross operator \cite{alaghi2013stochastic} with XOR as an absolute value subtractor and MUX as an adder.}
\label{fig:electrica_edge_detection}
\end{figure}

\begin{figure}[!h]
\centering
\captionsetup{justification=centering}
\includegraphics[width=\textwidth]{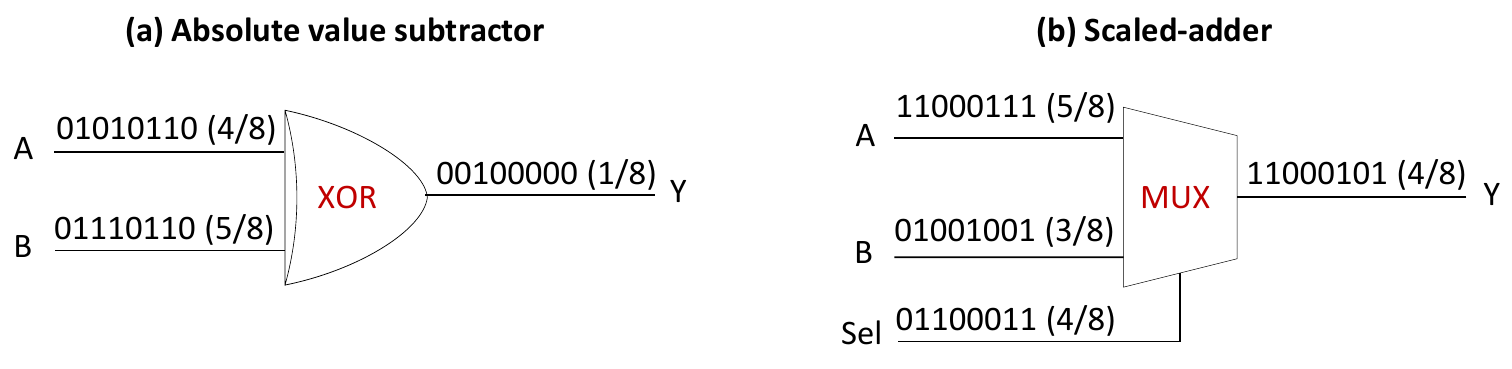}
\caption{(a) XOR gate as absolute value subtractor and (b) 2$\times$1 MUX as scaled adder.}
\label{fig:electrical_XOR_MUX}
\end{figure}

\begin{itemize}

\item\textbf{\textit{Absolute Value Subtractor:}} Figure~\ref{fig:electrical_XOR_MUX}(a) illustrates an XOR gate implementing a subtractor. This operation requires positively correlated bit streams with maximum overlap between '1's and '0's \cite{alaghi2013exploiting}. In the example, bit streams A=01010110 and B=01110110 are positively correlated with probability \textit{p\textsubscript{A}}=4/8 and \textit{p\textsubscript{B}}=5/8, respectively, which leads to \textit{p\textsubscript{Y}}=1/8. In general, the output of the XOR gate can be written as:

\begin{equation}
\label{eq:}
p_Y=   \begin{cases}
               p_A - p_B,\ \ \ \ &p_A > p_B\\
               p_B-p_A,\ \ \ \ &p_B > p_A
                 \end{cases}                 
\end{equation} 

which can be expressed as:

\begin{equation}
\label{eq:}
p_Y = \vert p_A-p_B\vert
\end{equation}

\item\textbf{\textit{Scaled-adder:}} This operation can be implemented using 2-1 MUX, as shown in Figure~\ref{fig:electrical_XOR_MUX}(b). The selection line has a probability of 1/2, which allows to downscale the output in order to keep the probability in the range [0,1]. While the bit streams to be added can be either uncorrelated or correlated\cite{alaghi2013survey}, the selection line needs to be uncorrelated with the inputs. The output of the MUX is given as:
\begin{equation}
\label{eq:}
p_Y = (1-p_{sel})p_A + p_{sel} p_B
\end{equation}  
since \textit{p\textsubscript{sel}}=1/2, the equation can be written as:
\begin{equation}
\label{eq:}
p_Y= \frac{1}{2} (p_A + p_B)
\end{equation}
The main drawback of this implementation is the reduced accuracy of the output due to downscaling the results by half. This can be overcome by doubling the \textit{BSL}, which, however, increases the latency. The design proposed in this work relies on cascaded MUXs, which induce precision loss but allow to maintain low hardware complexity. The impact of the precision loss on the application accuracy is evaluated, which allows to choose the most suitable \textit{BSL}.

\end{itemize}

A common issue in stochastic computing architectures is the overhead induced by SNGs. To overcome this issue, an adder allowing to reduce the number of LFSRs has been proposed in~\cite{budhwani2017taking}. The selection line of the MUX is connected to the Least Significant Bit (LSB) of the LFSR used to generate the MUX data inputs. The optical adder we propose relies on this efficient design. Since the same LFSR is used to generate correlated inputs~\cite{alaghi2013stochastic}, our design contains only a single LFSR to generate the bit streams for the XOR inputs and the selection lines of the MUXs.

\subsection{Optical Computing Architectures}

Integrated optics devices have proven their efficiency in the computing domain, among these devices are MZI, MRR and PhC devices. For instance, in~\cite{shen2017deep}, MZIs are used to design a fully optical neural network, which demonstrates 2 order of magnitude speedup, i.e., photodetection rate of 100GHz, compared to electronics implementation. In~\cite{perez2017toward}, MZI is used to design a reconfigurable mesh required to enable different functionalities in the architecture of microwave processors, such as FIR filters. In~\cite{li2013optical}, MRR is introduced in the design of optical lookup tables (OLUT), where Wavelength Division Multiplexing (WDM) allows executing multiple functions simultaneously. A Reconfigurable Directed Logic (RDL) architecture is designed using MRRs~\cite{xu2011reconfigurable}. It calculates the sum of products for a given function in two steps. First, the products are evaluated, then the sum of products is calculated. A 2-bit delayed XOR task is implemented on a 4$\times$4 swirl reservoir topology designed using nonlinear MRRs~\cite{denis2018all}. The results show that the design can reach 2.5$\times$10$^-4$ error rate. In~\cite{feldmann2020parallel}, the design of photonic hardware accelerator is proposed that can perform parallel matrix vector multiplication operations at a rate of several Tera Multiply-ACcumulate per second (TMAC/s), to process image using convolution filters. In~\cite{nozaki2012ultralow}, PhC cavity is proposed in the design of all-optical RAM, where writing, storage, reading, and erasing operations are demonstrated. In~\cite{yang2017ultracompact}, nanocavity is used in the implementation of all-optical logic gates using Kerr effect, such as NAND, XOR, and XNOR. An All-Optical-Gate (AOG) is designed in~\cite{constans2019iii} using PhC nanocavity, where light is used to control the transmission of light. Therefore, AOGs are essential in all-optical signal processing, where it is used to achieve all-optical sampling on chip. 

We investigated the combination of stochastic computing and integrated optics in~\cite{hassnaa-date-19}. We proposed the use of silicon photonics devices namely; MZI, MRR and all-optical add-drop filter, to implement an optical version of ReSC architecture~\cite{qian2010architecture}. The design can execute any arbitrary single input polynomial function and, in~\cite{el2020design}, we studied the impact of the \textit{BSL} (stochastic computing domain) and \textit{BER} (optical domain) on the application-level accuracy. 

In this work, we aim to use PhC nanocavities to design an all-optical stochastic architecture. We propose a transmission model to estimate the lasers power consumption and evaluate the computing accuracy. We investigate the design of XOR gate and MUX using nanocavities. We explore the device and system-level parameters in the design of cascaded gate architecture by implementing edge detection filter that relies on the proposed gates. 

\section{Photonics Crystal Nanocavity}
In this section, we introduce the PhC nanocavity device used to implement all-optical logic gates. The physical properties of the device and the implementation of an inverter are first detailed. Then, the design of XOR gate and MUX are presented. Finally, a transmission model of the nanocavity is proposed.

\subsection{Nanocavity Device Overview}

\begin{figure}[h]
\centering
\captionsetup{justification=centering}
\includegraphics[width = 0.6\textwidth]{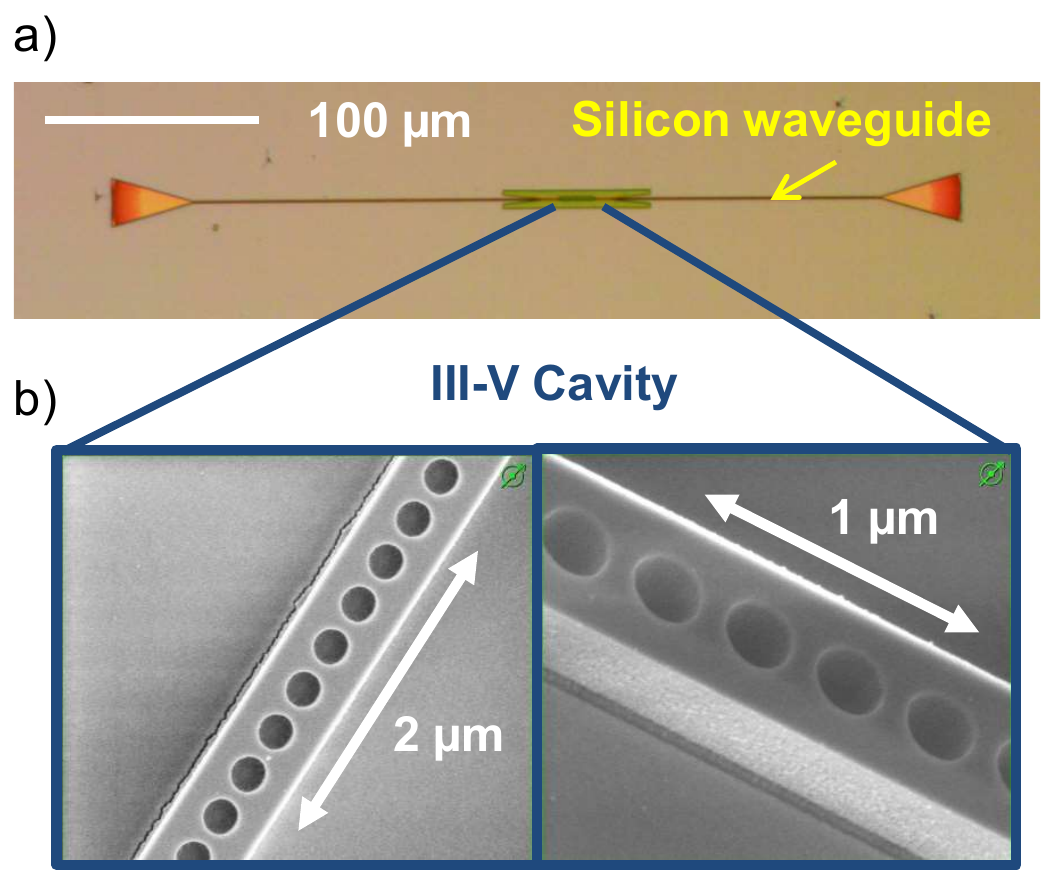}
\caption[Photographs of the studied PhC nanocavity.]{Photographs of the studied PhC nanocavity. a) A III-V semiconductor PhC cavity bonded on top of a silicon waveguide. b) Scanning electronic microscope top view photographies of III-V PhC cavities.}
\label{fig:PhC_photo}
\end{figure}

In this work, we use PhC nanocavity to implement all-optical logic gates. The structure is made of III-V semiconductor bonded on top of a silicon waveguide, as illustrated in Figure~\ref{fig:PhC_photo}(a). The PhC cavity itself consists of a waveguide drilled with holes (Figure~\ref{fig:PhC_photo}(b)). PhC nanocavity is a resonator that can act as a filter allowing only the resonant optical frequency to pass through. The implementation of fully optical gates using such cavity involves the triggering of nonlinear effect. This can be achieved using a high power optical signal to control the transmission of lower power optical signals. It has been shown that a fast (10ps) nonlinear response is possible with only about 100fJ of energy~\cite{husko2009ultrafast}, substantially outperforming MRRs~\cite{moille2016integrated}.

\subsection{All-optical NOT Gate}

\begin{figure}[!h]
\centering
\captionsetup{justification=centering}
\includegraphics[width = 0.85\textwidth]{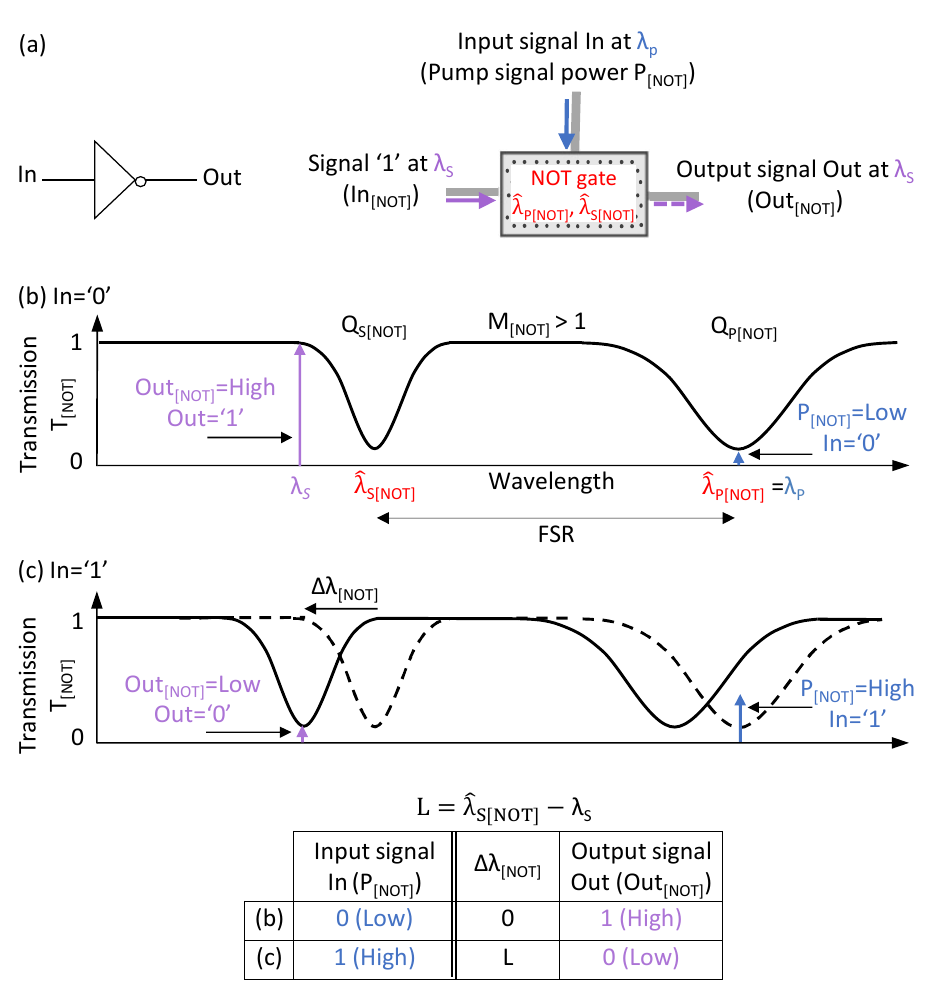}
\caption[An all-optical NOT gate implemented using nanocavity]{An all-optical NOT gate implemented using nanocavity: (a) logic gate and the equivalent nanocavity device representation, (b) gate transmission for logic input '0' and (c) gate transmission for logic input '1'.}

\label{fig:NOT_gate}
\end{figure}

As previously mentioned, the design of all-optical logic gates using nanocavity involves triggering nonlinear effects. We illustrate this principle using the implementation of all-optical NOT gate. As shown in Figure~\ref{fig:NOT_gate}(a), the NOT gate has an input \textit{In}, which corresponds to the pump signal injected into the nanocavity. The value of \textit{In} is given by its optical power \textit{P\textsubscript{[NOT]}} (i.e., low power means '0' and high power means '1'). Therefore, input signal \textit{In} controls the value of the output signal \textit{Out}, which corresponds to the output \textit{Out} of the NOT gate. The design of the nanocavity allows two (or more) resonances separated by Free Spectral Range (FSR). One resonance, in this case \textit{$\hat{\lambda}\textsubscript{P[NOT]}$}, is used to effectively inject a pump signal at \textit{${\lambda}\textsubscript{P}$}, which induces the spectral shift of the other resonances, i.e., $\hat{\lambda}_{S[NOT]}$. This modifies the transmission of the output signal at ${\lambda}_S$. The signal at ${\lambda}_S$ is always injected into the cavity as '1', as shown in Figure~\ref{fig:NOT_gate}(a). The operation of all-optical NOT gate is explained as follows:

\begin{itemize}[leftmargin=*]
\item \textit{In}='0' corresponds to $P_{[NOT]}$='Low' (Figure~\ref{fig:NOT_gate}(b)): in this case, the nanocavity is off-resonance, i.e., $\hat{\lambda}_{S[NOT]} \neq \lambda_{S}$. Thus, the transmission of the signal at $\lambda_{S}$ to the output is maximized, which leads to \textit{Out}='1'.

\item \textit{In}='1' corresponds to $P_{[NOT]}$='High' (Figure~\ref{fig:NOT_gate}(c)): The pump power detunes the resonance of the nanocavity by $\Delta\lambda_{[NOT]}$. The resonance of the cavity is then aligned to the output signal wavelength at $\lambda_{S}$, i.e., $\hat{\lambda}_{S[NOT]} = \lambda_{S}$. This leads to a strong attenuation of the signal and hence \textit{Out}='0'.
\end{itemize}

The fabrication process allows to control numerous parameters, such as $Q$ factors and resonance wavelengths. The design allows defining different $Q$ factors for each resonance, as shown in Figure~\ref{fig:NOT_gate}. Since we assume one pump and one output signals, it is possible to define $Q$ factors $Q_{P[NOT]}$ and $Q_{S[NOT]}$ at resonances $\hat{\lambda}_{P[NOT]}$ and $\hat{\lambda}_{S[NOT]}$, respectively. We define the ratio between $Q_{S[NOT]}$ and $Q_{P[NOT]}$ as the figure of merit ($M_{[NOT]}$) of the cavity ($M_{[NOT]}=Q_{S[NOT]}/Q_{P[NOT]}$). A nanocavity with a large figure of merits would allow to maintain efficient coupling of the pump signal power into the device, while significantly changing the transmission around the output signal wavelength. This would result in large gap between the cavity transmission for data '1' (i.e., no pump is applied) and data '0' (i.e., a pump signal is applied), i.e., high ER. The impact of the figure of merits is further discussed in Section~\ref{nanocavity_model}. In the sequel, we propose the implementation of all-optical XOR gate and MUX, which we use for the design of edge detection filter.

\subsection{Design of All-optical XOR Gate and MUX}
The design of an edge detection circuit requires XOR gate and MUX. The following introduces their implementation using nanocavity devices.

\begin{itemize}[leftmargin=*]

\begin{figure}[!t]
\centering
\captionsetup{justification=centering}
\includegraphics[width = 0.85\textwidth]{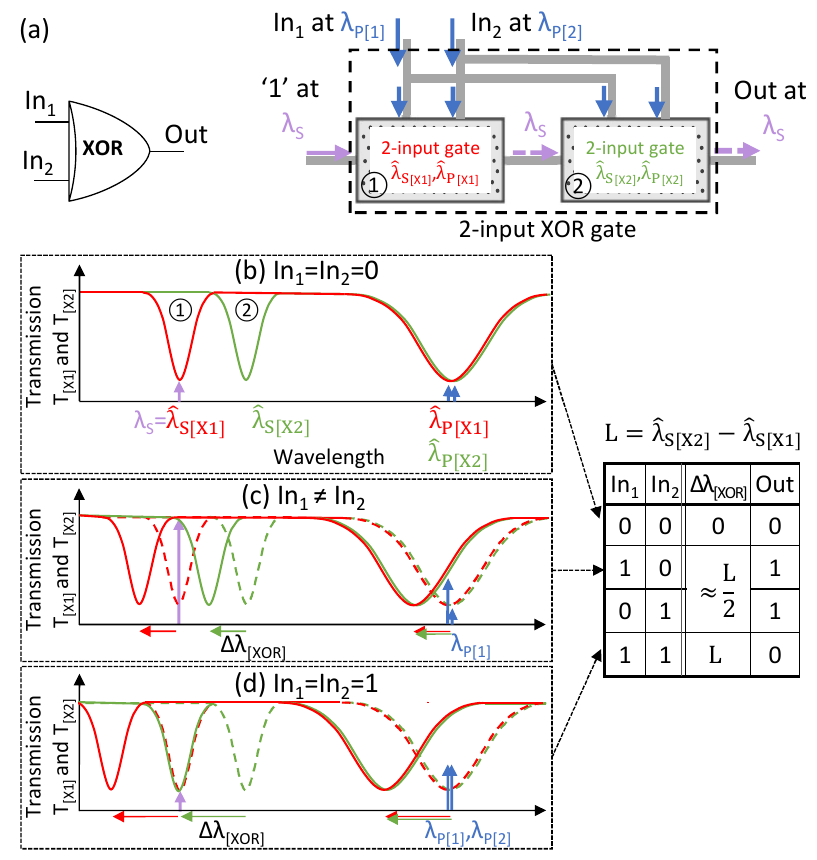}
\caption[Nanocavity operating as (a) a 2-input XOR gate]{Nanocavity operating as (a) a 2-input XOR gate implemented using two cascaded nanocavities. (b), (c), and (d) are the gate transmissions for different inputs scenarios.}
\label{fig:XOR_gate}
\end{figure}

\item\textbf{\textit{2-input XOR gate:}} A 2-input XOR gate is implemented using two cascaded nanocavities, as illustrated in Figure~\ref{fig:XOR_gate}(a). They are equal in $Q$ factors but different in the FSR. Nanocavities marked $\textcircled{1}$ and $\textcircled{2}$ resonate at $\hat{\lambda}_{S[X1]}$ and $\hat{\lambda}_{S[X2]}$, respectively. Inputs \textit{In\textsubscript{1}} and \textit{In\textsubscript{2}}, common for both cavities, are injected as pump signals into the cavities. The pump signals propagating at $\lambda_{p[1]}$ and $\lambda_{p[2]}$ are close in values to achieve the desired detuning. The signal at $\lambda_{S}$ is always '1'. It is tuned to match the resonance wavelength of the nanocavity marked $\textcircled{1}$ ($\hat{\lambda}_{S[X1]}=\lambda_{S}$) and hence initially, when no pump signal is injected (\textit{In\textsubscript{1}}=\textit{In\textsubscript{2}}=0), the signal is attenuated leading to \textit{Out}='0', as shown in Figure~\ref{fig:XOR_gate}(b). When one of the pump signals is high (i.e., \textit{In\textsubscript{1}} $\neq$ \textit{In\textsubscript{2}}), the resonance wavelengths of both cavities are shifted by $\Delta\lambda_{[XOR]}\approx 1/2 (\hat{\lambda}_{S[X2]}-\hat{\lambda}_{S[X1]})$. Since none of the resonance wavelengths is aligned with $\lambda_{S}$, this leads to the transmission of the signal at $\lambda_{S}$ with maximized power, i.e., \textit{Out}='1', as shown in Figure~\ref{fig:XOR_gate}(c). When the two pump signals are high (\textit{In\textsubscript{1}}=\textit{In\textsubscript{2}}='1'), as shown in Figure~\ref{fig:XOR_gate}(d), the resonance wavelengths of both cavities are detuned by $\Delta\lambda_{[XOR]}=(\hat{\lambda}_{S[X2]}-\hat{\lambda}_{S[X1]})$. Therefore, resonance wavelength $\hat{\lambda}_{S[X2]}$ is tuned to $\lambda_{S}$. Since $\hat{\lambda}_{S[X1]} \neq \lambda_{S}$, this leads to the transmission of the signal at $\lambda_{S}$ by the first device marked  $\textcircled{1}$ and to its attenuation by the second device marked  $\textcircled{2}$, hence \textit{Out}='0'.

\item\textbf{\textit{2$\times$1 MUX:}} A 2$\times$1 MUX is composed of a nanocavity resonating at $\hat{\lambda}_{S[MUX]}$ and controlled by the pump signal \textit{Sel}, as illustrated in Figure~\ref{fig:MUX_gate}(a). The pump signal allows selecting the input signal (i.e., \textit{In\textsubscript{1}} or \textit{In\textsubscript{2}}) to be transmitted to the output \textit{Out}. The selection is achieved by detuning the resonance of the nanocavity away from the required input signal. For this purpose, when no pump signal is injected (\textit{Sel}='0'), the resonance wavelength of the nanocavity is aligned with $\lambda_{S[1]}$, i.e., the wavelength of \textit{In\textsubscript{1}}, hence signal \textit{In\textsubscript{1}} is attenuated and signal \textit{In\textsubscript{2}} is transmitted to the output, as shown in Figure~\ref{fig:MUX_gate}(b), i.e., \textit{Out}=\textit{In\textsubscript{2}}. When a pump signal is injected (\textit{Sel}='1'), the nanocavity is detuned to $\lambda_{S[2]}$ ($\Delta\lambda_{[MUX]}= \lambda_{S[1]}-\lambda_{S[2]}$), thus leading to \textit{Out}=\textit{In\textsubscript{1}}, as illustrated in Figure~\ref{fig:MUX_gate}(c).
\begin{figure}[!h]
\centering
\captionsetup{justification=centering}
\includegraphics[width = 0.8\textwidth]{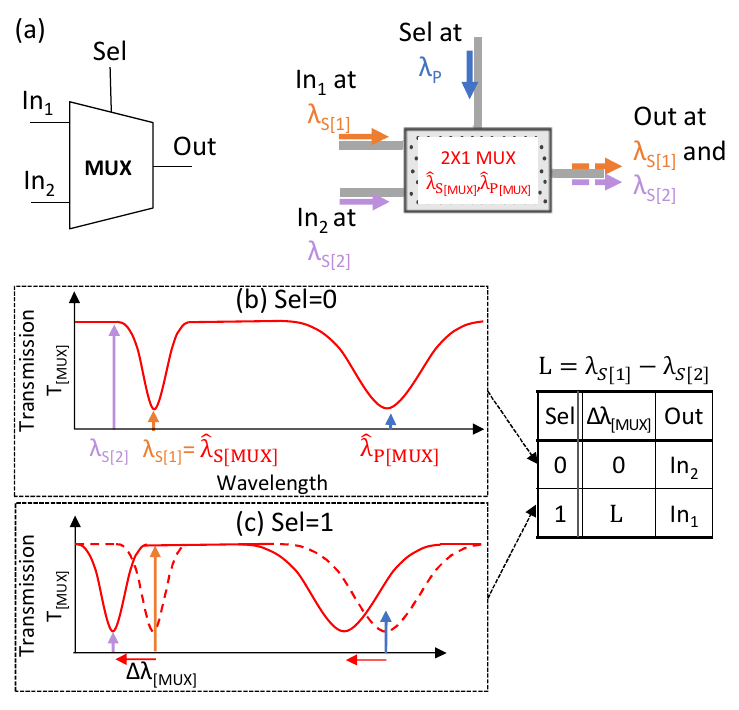}
\caption[Nanocavity operating as (a) a 2$\times$1 MUX]{Nanocavity operating as (a) a 2$\times$1 MUX. (b) and (c) MUX transmission.}
\label{fig:MUX_gate}
\end{figure}

The MUXs operate on multiple signals at different wavelengths and with multiple spacing. The nanocavities implementing MUXs thus need to be carefully defined, taking into account the resonant wavelength, the transmission bandwidth (i.e., $Q$ factor) and the detuning. 
In the following, we propose a model estimating the wavelength detuning and the transmission of a nanocavity, taking into account key device parameters and the applied pump power. 
\end{itemize}

\subsection{Nanocavity Model}
\label{nanocavity_model}
We propose a model allowing to design nanocavity based logic gates. The model allows i) estimating the wavelength detuning ($\Delta\lambda_{[gate]}$) according to the applied pump power ($P_{[gate]}$); and ii) the calculation of signal transmission ($T_{[gate]}$). Table~\ref{table:parameters_ch4} summarizes the device parameters, where $[gate]$ indicates the logic gate that is implemented using nanocavity, i.e., NOT, XOR, MUX, etc.

\begin{table}[!t]
\setlength{\tabcolsep}{1.5pt} 
\renewcommand{\arraystretch}{1.3} 
\small
\centering
\caption{Device parameters.}
\label{table:parameters_ch4}
\begin{tabular}{|c|c|c|}
\hline
\textbf{Parameter} & \textbf{Description}                                                                                                                      & \textbf{Unit} \\ \hline
$\hat{\lambda}_{P[gate]}$    & \begin{tabular}[c]{@{}c@{}}Resonance Wavelength around pump signal\\ (when no pump power is injected)\end{tabular}                        & nm            \\ \hline
$\hat{\lambda}_{S[gate]}$    & \begin{tabular}[c]{@{}c@{}}Resonance Wavelength around input signal \\ (when no pump power is injected)\end{tabular}                      & nm            \\ \hline
FSR                & Free spectral range (FSR=$\hat{\lambda}_{P[gate]}$-$\hat{\lambda}_{S[gate]}$)                                                                                                                & nm            \\ \hline
$Q_{P[gate]}$       & Quality factor around $\hat{\lambda}_{P[gate]}$                                                                                                                     & -             \\ \hline
$Q_{S[gate]}$       & Quality factor around $\hat{\lambda}_{S[gate]}$                                                                                                                      & -             \\ \hline
$M_{[gate]}$        & Figure of merit($M_{[gate]}$=$Q_{S[gate]}$/$Q_{P[gate]}$)                                                                                                                         & -             \\ \hline
$OTE_{[gate]}$      & \begin{tabular}[c]{@{}c@{}}Optical tuning efficiency (the detuning of the \\ nanocavity according to the applied pump power)\end{tabular} & -             \\ \hline
\end{tabular}%

\end{table}

\begin{figure}[!t]
\centering
\captionsetup{justification=centering}
\includegraphics[width = 0.75\textwidth]{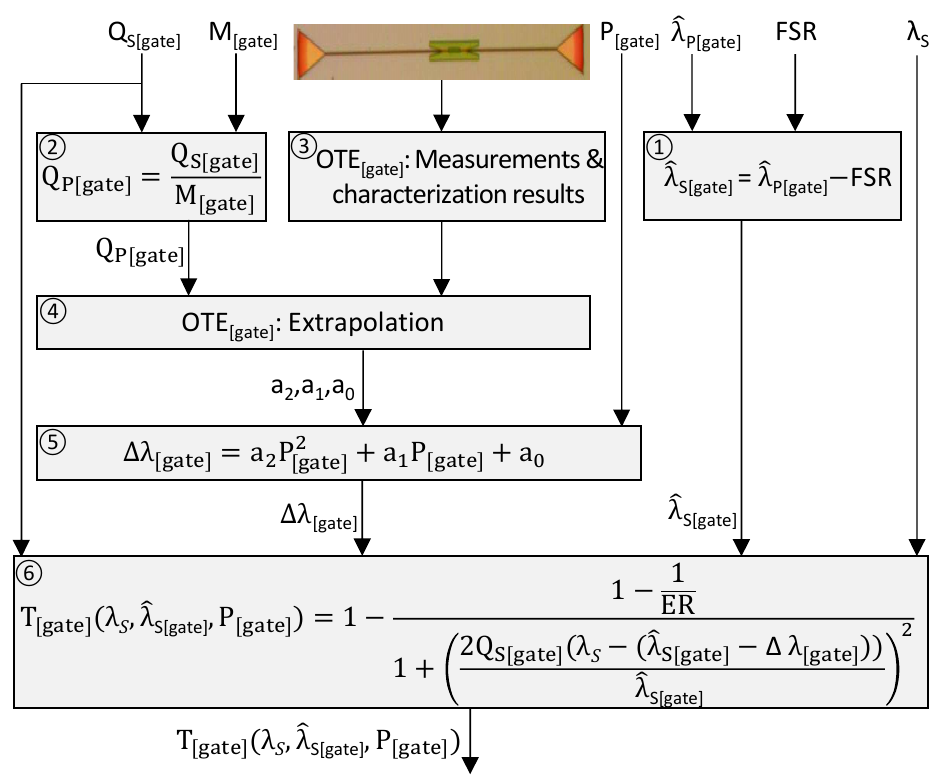}
\caption{Proposed model}

\label{fig:extrapolation_model}
\end{figure}

\begin{figure}[!b]
\centering

\captionsetup{justification=centering}
\includegraphics[width =\textwidth]{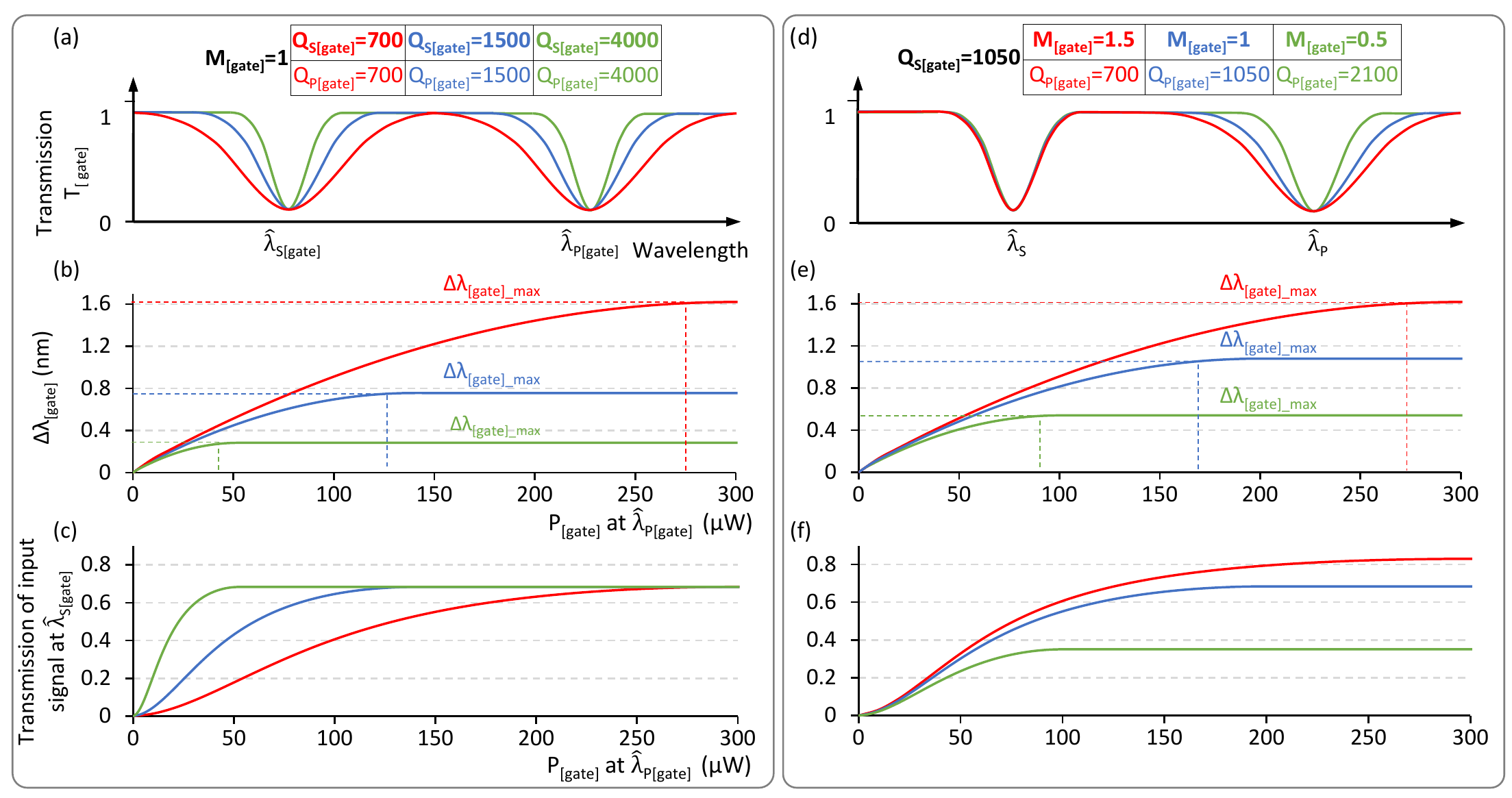}
\caption{(a) and (d) Transmission of nanocavity devices of ($Q_{S[gate]}$= 700, 1500, 4000, $M_{[gate]}$=1] and ($Q_{S[gate]}$=1050, $M_{[gate]}$=1.5, 1, 0.5). (b) and (e) The corresponding wavelength detuning ($\Delta\lambda_{[gate]}$) of (a) and (d), respectively. (c) and (f) The corresponding transmission of input signal according to the applied pump power.}
\label{fig:extra_scenario1}
\end{figure}

Inputs device parameters $\hat{\lambda}_{P[gate]}$ and FSR, shown in Figure~\ref{fig:extrapolation_model}, allow to evaluate $\hat{\lambda}_{S[gate]}$ (mark $\textcircled{1}$), when no pump power is applied. $Q_{P[gate]}$ (mark $\textcircled{2}$) is obtained from $Q_{S[gate]}$ and $M_{[gate]}$, which depend on the fabrication process and the cavity layout (e.g. width and length). The optical tuning efficiency ($OTE_{[gate]}$) is obtained through device characterizations (mark $\textcircled{3}$) and through linear extrapolation to a polynomial function (mark $\textcircled{4}$), which requires the targeted device parameters. The detuning (mark $\textcircled{5}$) is calculated by taking into account $Q_{P[gate]}$, the applied pump power ($P_{[gate]}$), and the $OTE_{[gate]}$. Finally, the transmission of the nanocavity is evaluated using Lorentzian approximation (mark $\textcircled{6}$) \cite{wu2015variation}.

We illustrate in Figure~\ref{fig:extra_scenario1} two scenarios using our model: i) different $Q_{S[gate]}$/same $M_{[gate]}$; and ii) same $Q_{S[gate]}$/different $M_{[gate]}$, respectively.

\begin{itemize}[leftmargin=*]
\item \textit{$\textbf{M\textsubscript{[gate]}=1}$} leads to the same $Q$ factor at pump and input signals resonances, as illustrated in Figure ~\ref{fig:extra_scenario1}(a) for $Q_{P[gate]}=Q_{S[gate]}$= 700, 1500, and 4000. The corresponding detuning ($\Delta\lambda_{[gate]}$) of the cavity is plotted for pump power ranging from 0 to 300$\mu$W, as shown in Figure~\ref{fig:extra_scenario1}(b). As it can be observed, the higher $Q_{P[gate]}$, the smaller the maximum detuning $\Delta\lambda_{[gate]\_max}$, which is due to the reduced coupling of the pump with the cavity. The transmission of the input signal at $\lambda_{S}$ according to the applied power is shown in Figure~\ref{fig:extra_scenario1}(c). While 70\% signal transmission can be obtained for all $Q_{S[gate]}$, the use of high $Q_{P[gate]}$ can lead to pump power reduction since the maximum transmission is reached earlier (50$\mu$W and 270$\mu$W for $Q_{S[gate]}$=4000 and $Q_{S[gate]}$=700, respectively).

\item \textit{$\textbf{M\textsubscript{[gate]} $\neq$ 1}$} leads to $Q_{P[gate]}$=700 and $Q_{P[gate]}$=2100 for $M_{[gate]}$=1.5 and $M_{[gate]}$=0.5, respectively, assuming $Q_{S[gate]}$=1050 (Figure~\ref{fig:extra_scenario1}(d)). As can be seen in Figure~\ref{fig:extra_scenario1}(f), the maximum signal transmission reaches 0.3 and 0.8 for $M_{[gate]}$=0.5 and $M_{[gate]}$=1.5, respectively. Reaching high \textit{ER} of the input signal is thus possible for high $M_{[gate]}$ figures, thus leading to opportunities to reduce the data signal power.

\end{itemize}
\section{Proposed Edge Detection Filter Architecture}
\label{edge_detection_filter}
In this section, we investigate the design of a stochastic filter application using photonic nanocavities. Detecting edges in an image can be implemented using first derivatives by sliding two dimensional filters over the pixels. The application of the filters involves subtracting and adding the input pixels with each other. In SC, absolute value subtraction and addition can be implemented using XOR gates and MUXs, respectively. The implementation of the gates in the optical domain has been discussed in the previous section. We then discuss the main design challenges related to computing accuracy and energy consumption.

\subsection{Architecture Overview}

The architecture we propose is generic and characterized by a size $N$. It is composed of one stage of 2\textit{\textsuperscript{N}} XOR gates (for the subtraction) followed by $N$ MUX stages (for the addition). Each MUX stage is composed of $2^N/2^n$ MUXs, where $n$ is the stage position in the addition tree ($1 \leqslant n \leqslant N$).

\begin{enumerate}[leftmargin=*]

\item\textbf{Design Patterns}

The architecture involves the following design patterns: 
\begin{itemize}[leftmargin=*]

\item Two XOR gates followed by a MUX allow implementing a sub-sum function. As illustrated in Figure~\ref{fig:main_edge_detection}(a), two input signals at $\lambda_{S[i]}$ and $\lambda_{S[i+1]}$ are injected into $XOR_{[i]}$ and $XOR_{[i+1]}$, respectively (mark $\textcircled{1}$ in the figure), where $i$ is the position of the XOR in the range $1 \leqslant i \leqslant 2^N$. For each gate, the transmission of the input signal to the output is controlled by a pump signal (mark $\textcircled{2}$) generated by an SNG (mark $\textcircled{3}$), as detailed later. The multiplexer $MUX_{[j_{1},1]}$ receives the signals transmitted through the XORs (mark $\textcircled{4}$), where $[j_{1},1]$ is the MUX at position $j_{1}$ in stage $n=1$ and $1 \leqslant j_{1} \leqslant 2^N/2$. Depending on the pump signal generated from SNG\textsubscript{5} (mark $\textcircled{5}$), the multiplexer either transmits the signal at $\lambda_{S[i]}$ or $\lambda_{S[i+1]}$.

\item Three MUXs allow implementing a sum function, as shown in Figure~\ref{fig:main_edge_detection}(b). The aim of the MUXs is to sum signals propagating at several wavelengths: a MUX at stage $n$ receives two sets of $2^n/2$ signals ($\textcircled{6}$ and $\textcircled{7}$) and outputs a single set of $2^n$ signals. For example, each input of the MUX at $n$=3 is composed of 4 signals wavelengths and its output is composed of 8 wavelengths. In this design, only one signal will propagate to the output, other signals will be filtered through the MUXs. However, the number of wavelengths that can potentially carry the signal increases with the MUX stage. This calls for a MUX design taking into account the number of signals to process and the distance between the wavelengths.
\end{itemize}
\begin{figure}[!t]
\centering
\captionsetup{justification=centering}
\includegraphics[width =\textwidth]{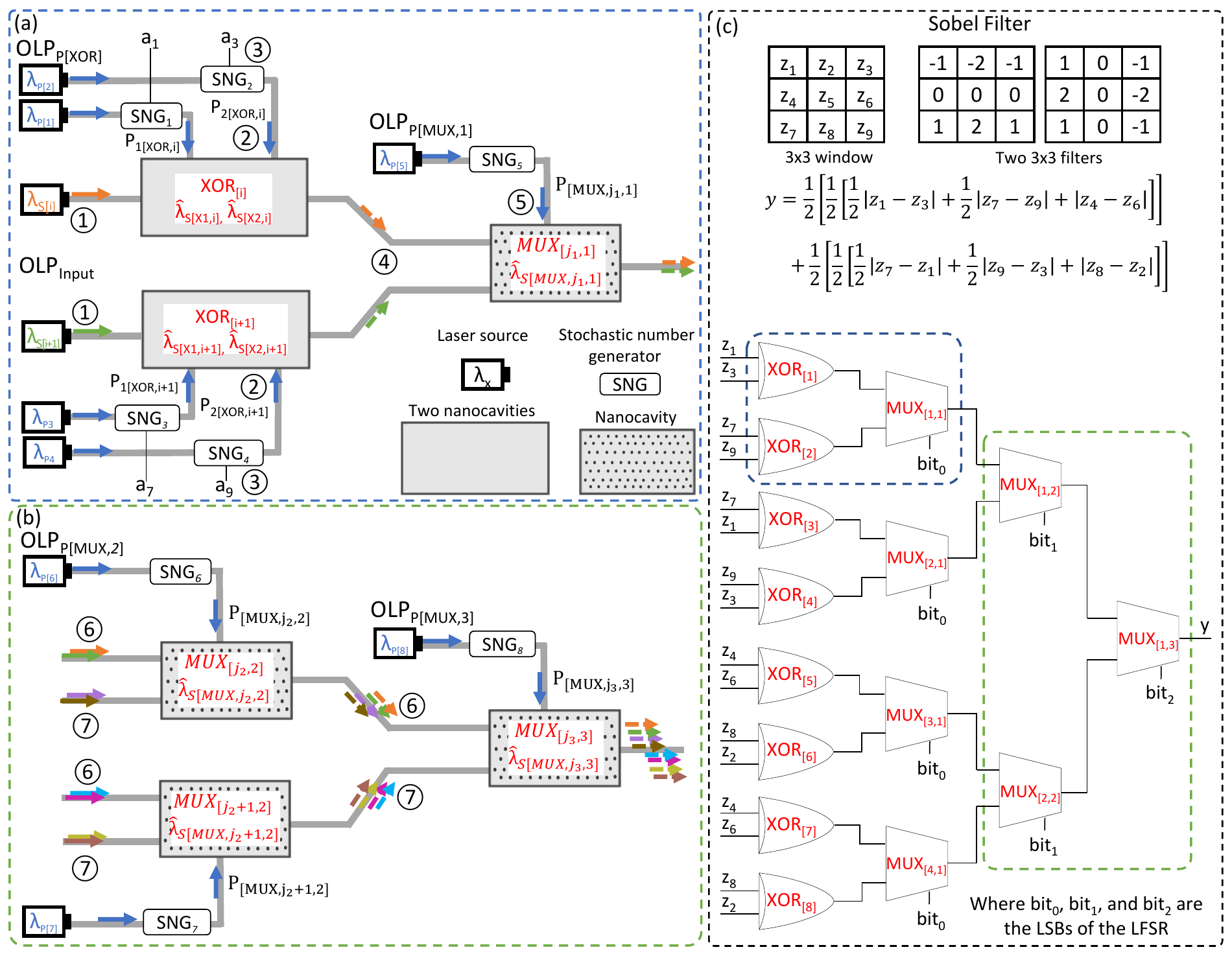}
\caption[The optical computing architecture of edge detection filters]{The The optical computing architecture of edge detection filters: a) proposed design pattern to implement subtraction and addition using XOR gates and MUXs; b) design pattern to implement a tree adder; c) architecture for the 3$\times$3 Sobel operator example.}

\label{fig:main_edge_detection}
\end{figure}

\item\textbf{Sobel Filter Architecture Example}

Figure~\ref{fig:main_edge_detection}(c) illustrates the design of a Sobel filter, where a 3$\times$3 window slides over the entire image to compute the gradient vector of the image. As shown in the figure, the design patterns are repeated through the entire architecture (see the blue and green dashed boxes). Each XOR receives two input pixels as pump signals, thus leading to a subtraction. The resulting signals propagate to the MUXs (that implement an adder-tree) and the output signal is transmitted to the photodetector. In order to keep the architecture symmetrical, we duplicate the input pixels for which coefficients 2 and -2 are applied in the Sobel filter. For instance, $XOR_{[7]}$ and $XOR_{[8]}$ are duplicated from $XOR_{[5]}$ and $XOR_{[6]}$, respectively. In optical domain, the design of the architecture requires i) eight lasers (i.e., one per XOR gate) emitting input signals at different wavelengths; and ii) 23 pump lasers (i.e., two per XOR gate and one per MUX). 

\item\textbf{Stochastic Number Generators (SNG)} 

The cavities are controlled by pump signals corresponding to stochastic numbers. As illustrated in Figure~\ref{fig:SNG_main_edge_detection}. Different SNGs are used for the XOR gates and the MUXs. However, in the proposed architecture, the same LFSR is used for the SNGs of all logic gates. The operation of the SNG according to the logic gates is detailed as follows:
\begin{figure}[!h]
\centering
\captionsetup{justification=centering}
\includegraphics[width =\textwidth]{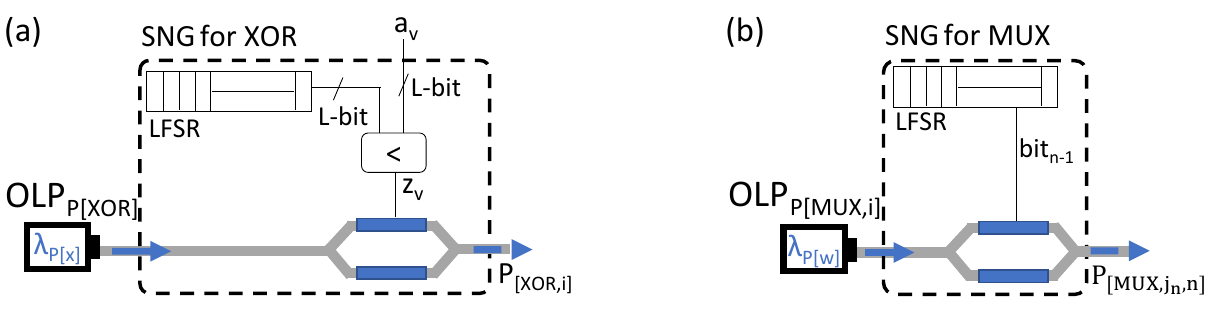}
\caption{SNGs for (a) XOR gates and (b) MUXs.}
\label{fig:SNG_main_edge_detection}
\end{figure}
\begin{itemize}[leftmargin=*]	
\item As shown in Figure~\ref{fig:SNG_main_edge_detection}(a), the XOR gates require electrical SNGs converting an input pixel ($a_v$) into a stochastic bit stream ($z_v$). For each XOR gate, $a_v$ is compared to the value generated by a LFSR: '1' is generated if the LFSR value is less than $a_v$; '0' is generated otherwise. The comparator controls a modulator, thus leading to the modulation of a signal continuously emitted by a laser at $\lambda_{P[x]}$. Bit '0' leads to a destructive interference in the modulator, hence a low pump signal is generated. Otherwise, a constructive interference in the modulator causes high pump signal ($P_{[XOR,i]}$) to be injected into the gate, thus allowing to implement the XOR function. In order to avoid crosstalk, each pump signal uses a dedicated wavelength. To generate correlated inputs, the same LFSR is used to generate the bit streams inputs for all XOR gates.

\item As shown in Figure~\ref{fig:SNG_main_edge_detection}(b), the selection line of the MUX only requires the generation of bit streams with the same number of zeros and ones (probability of 0.5) to generate $ P_{[MUX,j_n,n]}$ values. For this purpose, a modulator is directly controlled by a bit in the LFSR. In order to reduce the area and power overhead, the same LFSR (used for the XOR gates) is used to control several MUXs. This can be achieved without loss of accuracy by selecting bits at different positions.
\end{itemize}

\item\textbf{Transmission Spectrum and Device Characteristics}

As previously explained, the number of signals crossing the cavities increases with the stages. Figure~\ref{fig:trans_main_edge_detection} illustrates transmission examples corresponding to the architecture in Figure~\ref{fig:main_edge_detection}(c), where eight signals propagate using eight wavelengths. As detailed in the following, i) the distance between the wavelengths; and ii) the Q factor are key design parameters as they directly impact crosstalk and switching energy:
\begin{figure}[!t]
\centering
\captionsetup{justification=centering}
\includegraphics[width =0.8\textwidth]{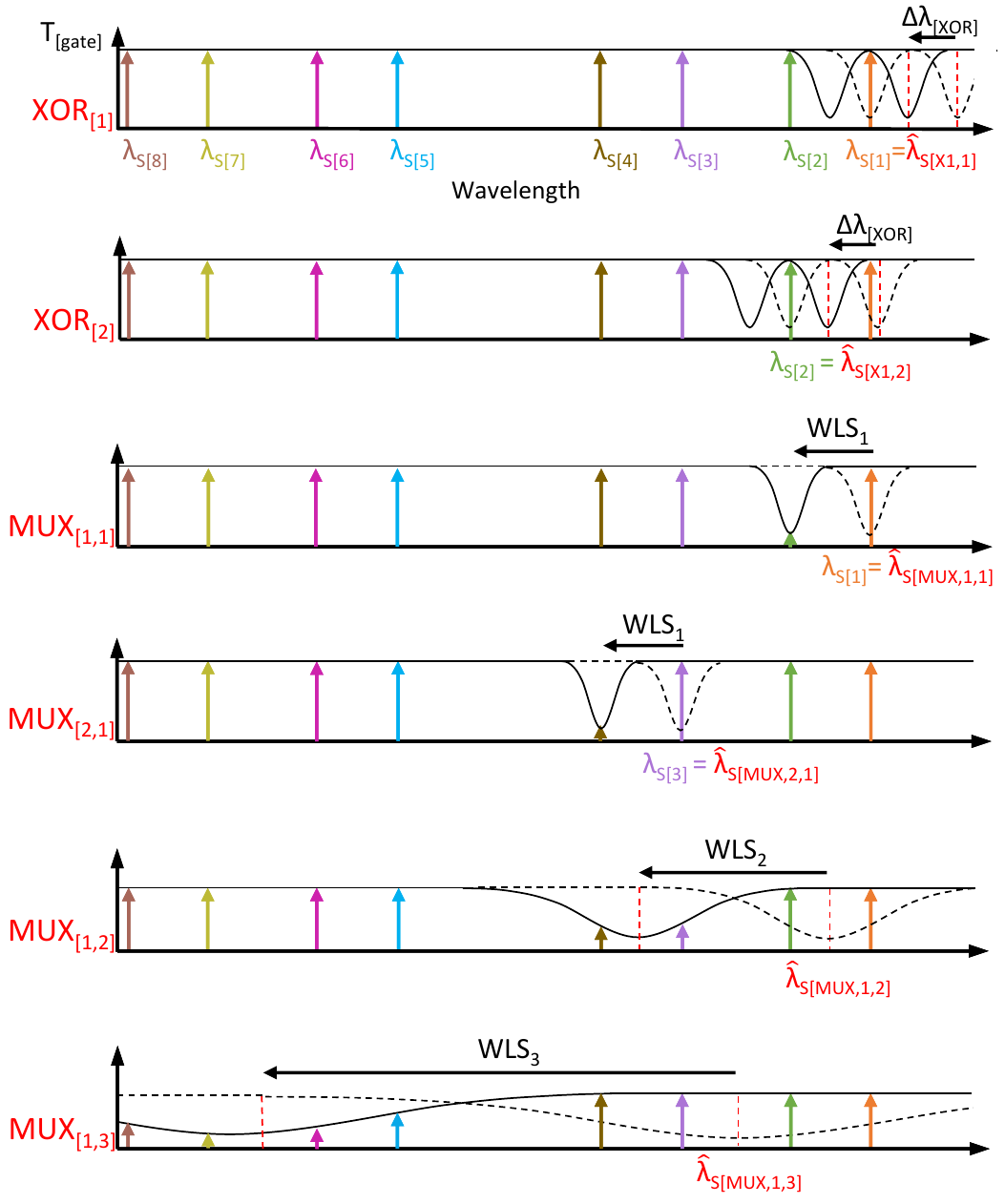}
\caption{The transmission of two XOR gates and one MUX per stage.}
\label{fig:trans_main_edge_detection}
\end{figure}
\begin{itemize}[leftmargin=*]
\item \textit{WLS\textsubscript{n}} corresponds to the wavelength spacing at stage $n$ of the MUX. The wavelengths are then regularly spaced following a hierarchy that suits the MUX tree. In the example, \textit{WLS\textsubscript{1}} is the distance between two consecutive signals in the first MUX stage, e.g., between $\lambda_{S[1]}$ and $\lambda_{S[2]}$, $\lambda_{S[3]}$ and $\lambda_{S[4]}$, etc. \textit{WLS\textsubscript{2}} is the distance between two consecutive sets of wavelengths in the second stage, e.g., between $\{\lambda_{S[1]}$,$\lambda_{S[2]}\}$ and $\{\lambda_{S[3]}$,$\lambda_{S[4]}\}$, $\{\lambda_{S[5]}$,$\lambda_{S[6]}\}$ and $\{\lambda_{S[7]}$,$\lambda_{S[8]}\}$, etc. 

\item $Q_{S[gate,n]}$ corresponds to the cavity $Q$ factor at stage $n$. Indeed, assuming the same $Q$ factor for all cavities in a stage allows using the same laser power per stage. Moreover, we assume both XOR gates and the MUXs in the first stage to have the same $Q$ factor. We define $Q_{S[XOR]}$, without $n$, as the $Q$ factor of the XOR gate around the input signal. Moreover, as the wavelength distance between signals to be multiplexed increases, the bandwidth of the cavity increases (i.e., $Q_{S[MUX,n]} > Q_{S[MUX,n+1]}$).
\end{itemize}
\end{enumerate}

To summarize, the design of the proposed architecture involves exploring numerous parameters, such as laser powers, wavelength distances and $Q$ factors. In the following, we further discuss their optimization according to computing accuracy and power consumption purposes.

\subsection{Design Challenges}
The design of such an architecture involves the optimization of computing accuracy, power consumption and processing time. The following summarizes key technological and system-level parameters we consider for the optimization of the architecture:

\begin{itemize}[leftmargin=*]
\item\textbf{BSL and BER:} computing accuracy depends on \textit{BSL} (stochastic domain specific) and \textit{BER} (optical domain specific). While both techniques result in power consumption, a reduction in the \textit{BER} should be preferred, since it can be achieved without impacting the processing time.

\item\textbf{Input signal power:} the architecture is composed of cascaded gates, which results in signal attenuation. In order to ensure a proper operation of the design, an input signal should be injected with a high enough optical power (typically 3$\mu$W to 10$\mu$W).

\item\textbf{Pump signal power:} it controls the wavelength detuning of the nanocavity and ranges from 100$\mu$W to 10mW scale. To prevent the input signal from detuning the cavity, we assume that its power should not exceed 10\% of the pump power.

\item\textbf{Wavelength spacing:} it impacts the power consumption as follows: small \textit{WLS} increases crosstalk and hence results in high \textit{BER}. This requires high lasers power for the input signals to overcome the crosstalk. On the contrary, larger \textit{WLS} contributes to a reduction in input signal power but calls for higher pump power to cover the larger wavelength detuning.
\end{itemize}

In the following, we present models allowing to explore these parameters.

\section{Implementation and Model}
In this section, we present an analytical model to evaluate the error induced from the stochastic computing technique and the optical transmission. Moreover, we develop a transmission model for the edge detection filter to estimate the power consumption. We also define the required design parameters and an exploration methodology.

\subsection{Error Evaluation}
Two types of errors are considered: i) errors related to stochastic computing domain; and ii) errors related to optical domain as discussed in the following:

\begin{itemize}[leftmargin=*]
\item\textbf{\textit{ED\textsubscript{BSL}}}: an error distance induced by the approximation when generating stochastic bit streams. This error is defined as:
\begin{equation}
\label{eq:bsl_error}
ED_{BSL} = \vert \acute Y -Y \vert
\end{equation}
\noindent where $Y$ is the error free result and $\acute{Y}$ is the approximated result for a given \textit{BSL}.

\item\textbf{\textit{ED\textsubscript{Trans}}}: an error distance induced by the optical transmission and occurs at the photodetector side. It is indicated by the \textit{BER}, i.e., the ratio of incorrectly transmitted bits. \textit{ED\textsubscript{Trans}} is given as:
\begin{equation}
\label{eq:trans_error}
ED_{Trans} = \vert \acute{\acute Y} - \acute Y \vert
\end{equation}
\noindent where $\acute{\acute{Y}}$ is the approximated result considering given $BSL$ (related to $\acute{Y}$) and \textit{BER}. This error can be enhanced using high laser power. As a result, the total error (worst-case error) can be defined as:
\begin{equation}
\label{eq:total_error}
ED_{Total} = ED_{BSL} + ED_{Trans}
\end{equation}

\noindent We use PSNR as a metric to evaluate the computing accuracy when processing an image as follows:
\begin{equation}
\label{eq:PSNR}
PSNR_{Total} = 10 \times log_{10} \big(\frac{MAX_{I}^{2}}{MSE_{Total}} \big)   
\end{equation}
\noindent where \textit{MAX\textsubscript{I}} is the maximum pixel in the error free image defined as 255 for 8-bit pixels. \textit{MSE\textsubscript{Total}} is the Mean Square Error given as:
\begin{equation}
\label{eq:MSE}
MSE_{Total} = \frac{1}{M\times K} \sum^{M}_{i=1} \sum_{j=1}^{K} ED_{Total}(i,j)^2
\end{equation}

\noindent where \textit{M} and \textit{K} are the number of rows and columns in the image, respectively. \textit{ED\textsubscript{Total(i,j)}} is the total error distance from processing a pixel at position \textit{(i,j)} in the image.

\end{itemize}

\subsection{Edge Detection Transmission Model}
In order to estimate the \textit{BER} of the architecture, we need to define the transmission of the signals. As defined in Section~\ref{edge_detection_filter}, an edge detection architecture of size $N$ is composed of 2\textit{\textsuperscript{N}} XOR gates, where each gate is designed using two nanocavities connected in series. Each XOR gate transmits one of $2^N$ input signals through $N$ MUXs. The transmission ($T_{[i]}$) of input signal $i$, propagating at $\lambda_{S[i]}$ through two nanocavities of the XOR gate and $N$ MUXs is given as: 
\begin{equation}
\label{eq:tran_signal}
\begin{split}
T_{[i]} =\ \  & \underbrace{T_{[X1]}(\lambda_{S[i]},\hat\lambda_{S[X1,i]},P_{\{1,2\}[XOR,i]})}_{\text{Transmission through the first cavity in XOR gate}} \times \\ 
& \underbrace{T_{[X2]}(\lambda_{S[i]},\hat\lambda_{S[X2,i]},P_{\{1,2\}[XOR,i]})}_{\text{Transmission through the second cavity in XOR gate}} \times \\
&  \underbrace{\prod_{n=1}^{N}T_{[MUX]}(\lambda_{S[i]},\hat\lambda_{S[MUX,j_{n},n]},P_{[MUX,j_{n},n])}}_{\text{Transmission through N MUXs}}
\end{split}
\end{equation}

\noindent where $j_{n}=\lceil i/2^n\rceil$ is the MUX position in stage $n$ and $1 \leqslant j_{n} \leqslant 2^{N}/2^{n}$.

\noindent From the signal transmission, \textit{SNR} is calculated as follows:
\begin{equation}
\label{eq:SNR_ch4}
SNR = OLP_{Input} \times \frac{R}{I} \times \Big( T_{[i]}-\sum^{M}_{\substack{k=1\\
                   k\neq i}} T_{[K]}\Big)
\end{equation}
\noindent where \textit{OLP\textsubscript{Input}} is the laser power of input signal at $\lambda_{S[i]}$ injected into the XOR gate. $R$ and $I$ are the photodetector responsivity and internal noise, respectively. $T_{[i]}$, in this case, is the transmission of signal \textit{i} as '1', while the other crosstalk signals $k$ are transmitted as '0'. $T_{[k]}$ is the transmission of the crosstalk signals $k$ as '1' while signal i is transmitted as '0', where $M=2^N$. The $BER$ assuming ON/OFF Key ($OOK$) modulation of the input signals is given by: 
\begin{equation}
\label{eq:BER_ch4}
BER= \frac{1}{2}erfc \big(\frac{SNR}{2\sqrt{2}}\big)
\end{equation}

\subsection{Nanocavity Design Parameters}
\label{nanocavity_design_parameters}
The evaluation of $T_{[i]}$ depends on $\lambda_{S[i]}$, $\hat{\lambda}_{S[gate]}$, and $P_{[gate]}$ parameters, which we define according to the methodology detailed in the following:

\begin{itemize}[leftmargin=*]
\item\textbf{Signal Wavelengths, Cavity Resonances and Spacing:} As previously explained, \textit{WLS\textsubscript{{n}}} corresponds to the shifting distance of the cavities located in stage $n$. Based on Figure~\ref{fig:trans_main_edge_detection}, we assume \textit{WLS\textsubscript{{3}}} $>$ \textit{WLS\textsubscript{{2}}} $>$ \textit{WLS\textsubscript{{1}}}. In the XOR stage, each gate will operate on a signal propagating at $\lambda_{S[i]}$, where $i$ is the row input number ($1 \leqslant i \leqslant 2N$). We set to 1542nm the baseline wavelength $\lambda_{S[1]}$ (i.e., the first input signal in Figure~\ref{fig:main_edge_detection}(c)). The subsequent signal wavelengths are assigned as follows:
\begin{equation}
\label{eq:lambda_input}
\lambda_{S[i]} = \lambda_{S[1]} - \sum^{N}_{n=1}\Big(\lfloor \frac{i-1}{2^n-1}\rfloor mod 2\Big) \times WLS_{n}
\end{equation}
\noindent For each XOR gate, we set the first and second resonance (i.e., $\hat{\lambda}_{S[X1,i]}$ and $\hat{\lambda}_{S[X2,i]}$) according to the signal wavelength $\lambda_{S[i]}$ and the assumed detuning $\Delta\lambda_{[XOR]}$: 
\begin{equation}
\label{eq:res_XOR1}
\hat\lambda_{S[X1,i]} = \lambda_{S[i]}
\end{equation}
\begin{equation}
\label{eq:res_XOR2}
\hat\lambda_{S[X2,i]}=\hat\lambda_{S[X1,i]}+ \Delta\lambda_{[XOR]}
\end{equation}
\noindent The resonance at rest of each MUX is defined by the mean wavelength of the first set of input signals:
\begin{equation}
\label{eq:res_MUX}
\hat\lambda_{S[MUX,j_{n},n]}=\frac{\lambda_{S[2^n(i-1_+1]}+\lambda_{S[2^n(i-1)+2^{n-1}]}}{2}
\end{equation}
\noindent where  $j_{n}= \lceil i/2^{n} \rceil$ is the MUX position in stage $n$.

\item\textbf{Pump Power:} we assume the same pump lasers power (\textit{OLP\textsubscript{P}}) injected into the cavities located in the same stage. The pump powers received by XOR gates are defined by:
\begin{equation}
\label{eq:pump_XOR}
P_{\{1,2\}[XOR,i]} = 
\begin{cases}
OLP_{P[XOR]}\times IL,  &z_{v}=1\\
OLP_{P[XOR]} \times IL \times ER,  & z_{v} =0\\
\end{cases}
\end{equation}
\noindent where where \textit{IL} is the Insertion Loss, \textit{ER} is the Extinction Ratio, $z_{v}$ is the bit streams of the input pixels for XOR gate. The pump powers received by the MUXs are given as:
\begin{equation}
\label{eq:pump_MUX}
P_{[MUX,j_{n},n]} = 
\begin{cases}
OLP_{P[MUX,n]}\times IL,  & LFSR \ bit_{n-1} = 1\\
OLP_{P[MUX,n]} \times IL \times ER,  & LFSR \ bit_{n-1} = 0\\
\end{cases}
\end{equation}
\noindent To ensure that the input power signal does not contribute to the detuning of the nanocavity, we set the maximum power of the input signal to 10\% of the cavity pump power.

\item\textbf{Algorithm:} the following summarizes the steps we follow to explore the design space:
\begin{enumerate}
\item Define input parameters: figure of merits ($M_{[gate]}$), wavelength of input signal ($\lambda_{S[1]}$), and targeted \textit{BER} at the photodetector. 
\item From the experimental results, use $Q_{S[gate]}$, $Q_{P[gate]}$, $\hat{\lambda}_{S[gate]}$ and $\hat{\lambda}_{P[gate]}$ to calibrate the PhC nanocavity model. Validate that the transmissions model and measurements are well correlated. 
\item For XOR gate design, explore $\Delta\lambda_{[XOR]}$ and $Q_{s[XOR]}$ to minimize laser power. This requires setting the resonance wavelengths of the XOR gate; $\hat{\lambda}_{S[X1,1]}$ and $\hat{\lambda}_{S[X2,1]}$ according to Equations~\ref{eq:res_XOR1} and~\ref{eq:res_XOR2}, respectively. 
\item For the MUX design, iterate from stage 1 to $N$ to:
\begin{enumerate}[label=(\alph*)]	
\item Set the resonance wavelength of the \textit{MUX\textsubscript{[1,1]}} to $\lambda_{S[1]}$ (Equation~\ref{eq:res_MUX}).
\item Explore \textit{WLS\textsubscript{1}} and $Q_{s[MUX,1]}$ to minimize \textit{BER} at the output stage, and select the desired \textit{BER}. This allows defining $\lambda_{S[2]}$ according to Equation~\ref{eq:lambda_input} and the resonance wavelength of \textit{XOR\textsubscript{[2]}} and \textit{MUX\textsubscript{[1,2]}} according to Equations~\ref{eq:res_XOR1} -~\ref{eq:res_MUX}.
\item Repeat step 4.b to explore \textit{WLS\textsubscript{2}} and $Q_{s[MUX,2]}$. By selecting a \textit{BER}, $\lambda_{S[3]}$ and $\lambda_{S[4]}$ are now evaluated using the corresponding \textit{WLS\textsubscript{2}} (Equation~\ref{eq:lambda_input}). Accordingly, the resonance wavelengths of \textit{XOR\textsubscript{[3]}}, \textit{XOR\textsubscript{[4]}}, and \textit{MUX\textsubscript{[2,1]}} are defined (Equations~\ref{eq:res_XOR1} -~\ref{eq:res_MUX}).
\item Repeat step 4.b again for the next stage until stage $N$. At this point, all \textit{WLS} are defined. This allows calculating the wavelengths of the rest of input signals and the resonance wavelengths of the remaining devices.
	\end{enumerate}	
\item According to the input lasers power and pump lasers power, estimate the energy per bit (Equations~\ref{eq:pump_XOR} and~\ref{eq:pump_MUX}).
\item Process an image and evaluate the application-level computing accuracy for a given \textit{BSL} and input lasers power (Equations~\ref{eq:bsl_error} -~\ref{eq:PSNR}).
\end{enumerate}	

\end{itemize}

\section{Results} 

In this section, we target a NOT gate of a given Q factor and compare the transmission and detuning using our proposed model and the experimental characteristics. We evaluate the lasers powers for a NOT gate and present the valid range of wavelength detuning. We introduce the design of XOR gate and MUX by exploring the design space in each stage. We process an image using the proposed architecture and we evaluate the computing accuracy, energy consumption and processing time.

\subsection{Model Calibration}
\begin{figure}[!h]
\centering
\captionsetup{justification=centering}
\includegraphics[width = 0.7\textwidth]{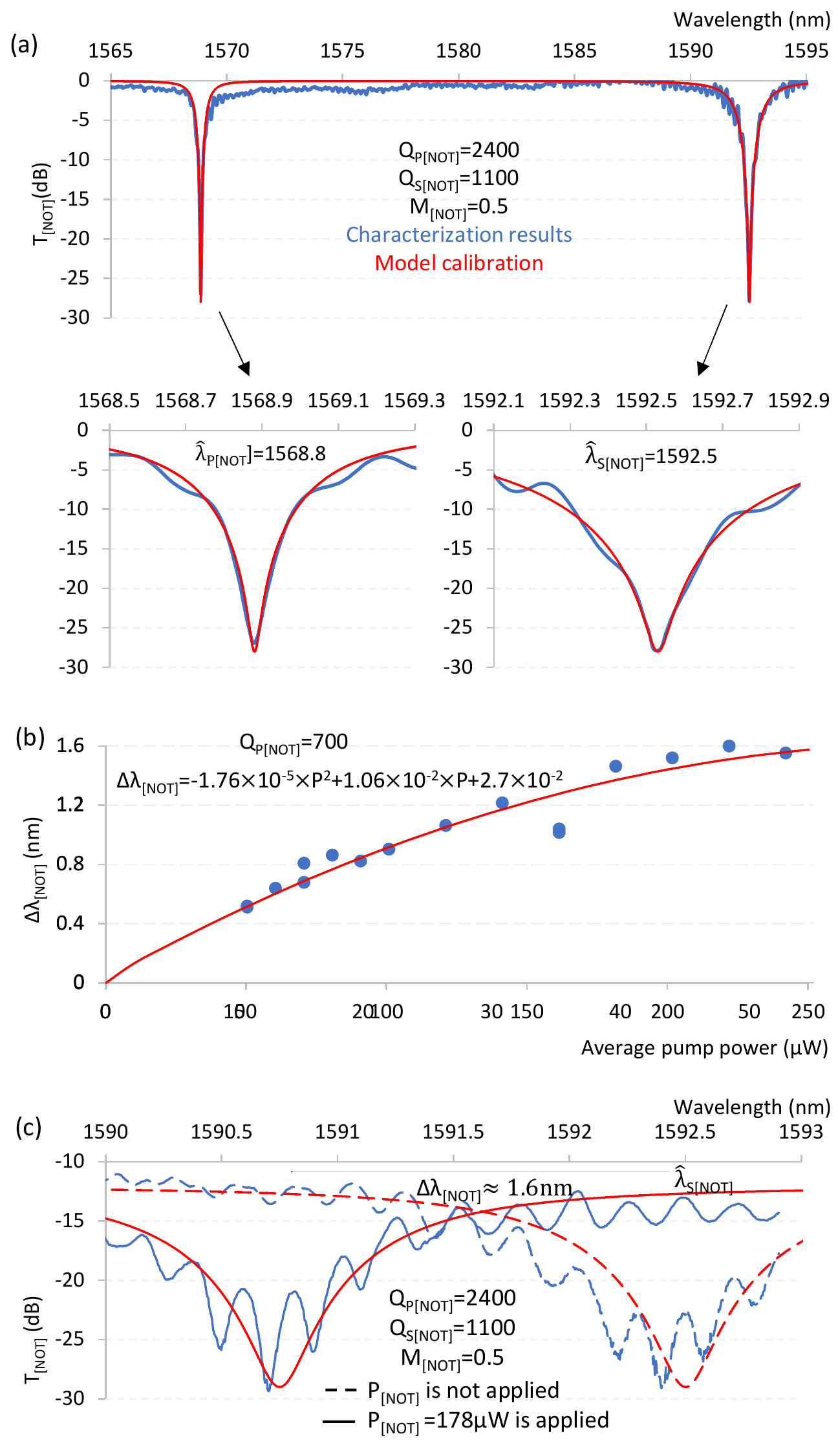}
\caption[Characterization results and model calibration]{Characterization results and model calibration: (a) Transmission when no pump power is applied, (b) wavelength detuning according to the average pump power, and (c) transmission when a pump power is injected.}
\label{fig:model_calibration}
\end{figure}

In the following, we detail the model calibration according to the experimental results for a NOT gate. As it can be observed from the transmission results reported in Figure~\ref{fig:model_calibration}(a), the gate is characterized by resonance wavelengths at $\hat{\lambda}_{S[NOT]}$=1592.5nm (around input signal) and $\hat{\lambda}_{P[NOT]}$=1568.8nm (around pump signal), which leads to FSR=24nm. At $\hat{\lambda}_{S[NOT]}$ and $\hat{\lambda}_{P[NOT]}$ resonances, the 3dB bandwidth of the nanocavity is 1.44nm and 0.65nm, respectively, which induces $M_{[NOT]}$=0.5. We calibrate the model using these parameters and, as it can be seen in the figure, a good correlation is obtained.

\begin{figure}[!t]
\centering
\captionsetup{justification=centering}
\includegraphics[width = 0.85\textwidth]{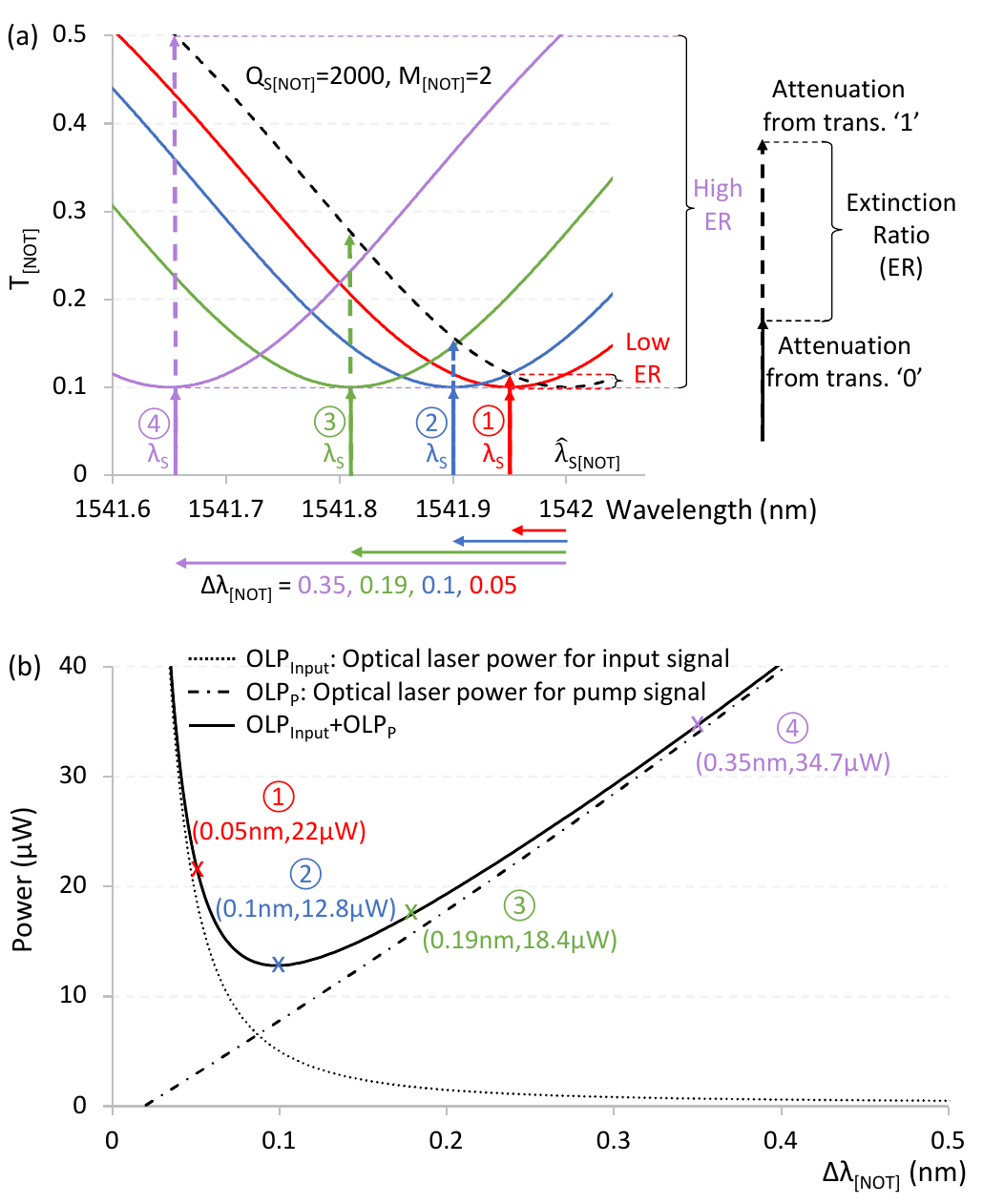}
\caption[For a nanocavity of $Q_{S[NOT]}=2000$ and $M_{[NOT]}=2$]{For a nanocavity of $Q_{S[NOT]}=2000$ and $M_{[NOT]}=2$: (a) The transmission assuming $\Delta\lambda_{[NOT]}=0.05$, 0.1, 0.19, and 0.35nm. (b) Lasers powers according to $\Delta\lambda_{[NOT]}$ ranges from 0 to 0.5nm.}

\label{fig:Not2_result}
\end{figure}

Figure~\ref{fig:model_calibration}(b) shows the measured nonlinear cavity detuning ($\Delta\lambda_{[NOT]}$) corresponding to an off-chip pump average power ranging from 0 to 250$\mu$W. This corresponds to an on-chip pump pulse energy up to 800fJ, for a cavity \textit{Q} factor=700. These pulsed mode measurements could be extrapolated to quasi CW excitation, which will be considered here assuming pulse duration is equal to the carrier lifetime, about 10ps. Thus, the off-chip average pump level 250$\mu$W corresponds to on-chip peak power roughly equals to 100mW.  Depending on the $Q$ factor and the material used, these numbers might change. In fact, the resonator here has been designed for maximized speed, hence low $Q$, trading off with energy efficiency. A different balance would target an order of magnitude larger $Q$. Figure~\ref{fig:model_calibration}(c) illustrates the transmission of the cavity at $\hat{\lambda}_{S[NOT]}$ under a 78mW on-chip peak power (178$\mu$W pump power). This leads to around 1.6nm blue shift of the resonance, which we observe for both measurement and model, thus validating the calibration.

In the following, we explore the impact of the signal detuning ($\Delta\lambda_{[NOT]}=\hat{\lambda}_{S[NOT]}-\lambda_{S}$) on the lasers powers, where $\hat{\lambda}_{S[NOT]}$ is the cavity resonance at rest. We consider a nanocavity with $Q_{S[NOT]}$=2000, $M_{[NOT]}$=2 and $\hat{\lambda}_{S[NOT]}$=1542nm. In Figure~\ref{fig:Not2_result}(a), we assume transmission scenarios for $\Delta\lambda_{[NOT]}$=0.05nm, 0.1nm, 0.19nm, and 0.35nm. Two optical signals are injected: $OLP_{Input}$ and $OLP_P$ correspond to the optical power of input signal and pump signal, respectively. As illustrated in Figure~\ref{fig:Not2_result}(a), $\Delta\lambda_{[NOT]}$=0.05 (mark $\textcircled{1}$) requires the lowest $OLP_P$ value due to the small shift in the resonant wavelength. On the other hand, this results in a rather low 0.7dB $ER$, which is compensated by using a high $OLP_{Input}$ value. Higher $\Delta\lambda_{[NOT]}$, such as 0.1nm (mark $\textcircled{2}$), 0.19nm (mark $\textcircled{3}$), and 0.35nm (mark $\textcircled{4}$), leads to an increase in the ER=1.7dB, 4.3dB, and 6.9dB, respectively. This contributes to lower $OLP_{Input}$ but induces higher $OLP_P$ due to the larger wavelength detuning distance.

To further explore the design space, we investigate the design power consumption by considering lasers powers, i.e., $OLP_{Input}$ and $OLP_{P}$. We assume $BER=10^{-1}$ and $\Delta\lambda_{[NOT]}$ ranging from 0 to 0.5nm. We define the valid range when $OLP_{Input}$ accounts for 10\% or less of $OLP_{P}$. As it can be seen in Figure~\ref{fig:Not2_result}(b), the power consumption is dominated by $OLP_{Input}$ for $\Delta\lambda_{[NOT]}<0.1$nm. At $\Delta\lambda_{[NOT]}$=0.05nm (mark $\textcircled{1}$), we obtain $OLP_{P}$=2.9$\mu$W and $OLP_{Input}$=19.1$\mu$W (for a total power of 22$\mu$W). This implies an input signal power (injected by $OPL_{Input}$) exceeding 10\% of the pump signal power (injected by $OPL_{P}$). Therefore, $\Delta\lambda_{[NOT]}$=0.05nm is an invalid option. Although $\Delta\lambda_{[NOT]}$=0.1nm (mark $\textcircled{2}$) leads to optimal total power consumption, it is not a valid design option, since the $OLP_{Input}$ accounts for 39\% of the total power received by the cavity. From $\Delta\lambda_{[NOT]}$=0.19nm (mark $\textcircled{3}$) to $\Delta\lambda_{[NOT]\_max}$=1.13nm, the design becomes valid but leads to power overhead. Hence the power is dominated by $OLP_{P}$ due to the large wavelength distance needed to reach the input signal. For example, $\Delta\lambda_{[NOT]}$=0.35nm (mark $\textcircled{4}$) involves $OLP_{P}$=33.9$\mu$W and $OLP_{Input}$=0.7$\mu$W, which increases the power consumption by 2.7$\times$ compared to the optimal $\Delta\lambda_{[NOT]}$. Each nanocavity of a given $Q_{S[NOT]}$ has a unique range of wavelength detuning that varies between 0 and $\Delta\lambda_{[NOT]\_max}$. However, the minimum detuning is specified according to the ratio of the injected input power to the pump power signals. In the sequel, we explore the power consumption in the design of $XOR$ gates considering nanocavities of different $Q$ factors.

\subsection{Design of XOR Gate}
\begin{figure}[!h]
\centering
\captionsetup{justification=centering}
\includegraphics[width = 0.85\textwidth]{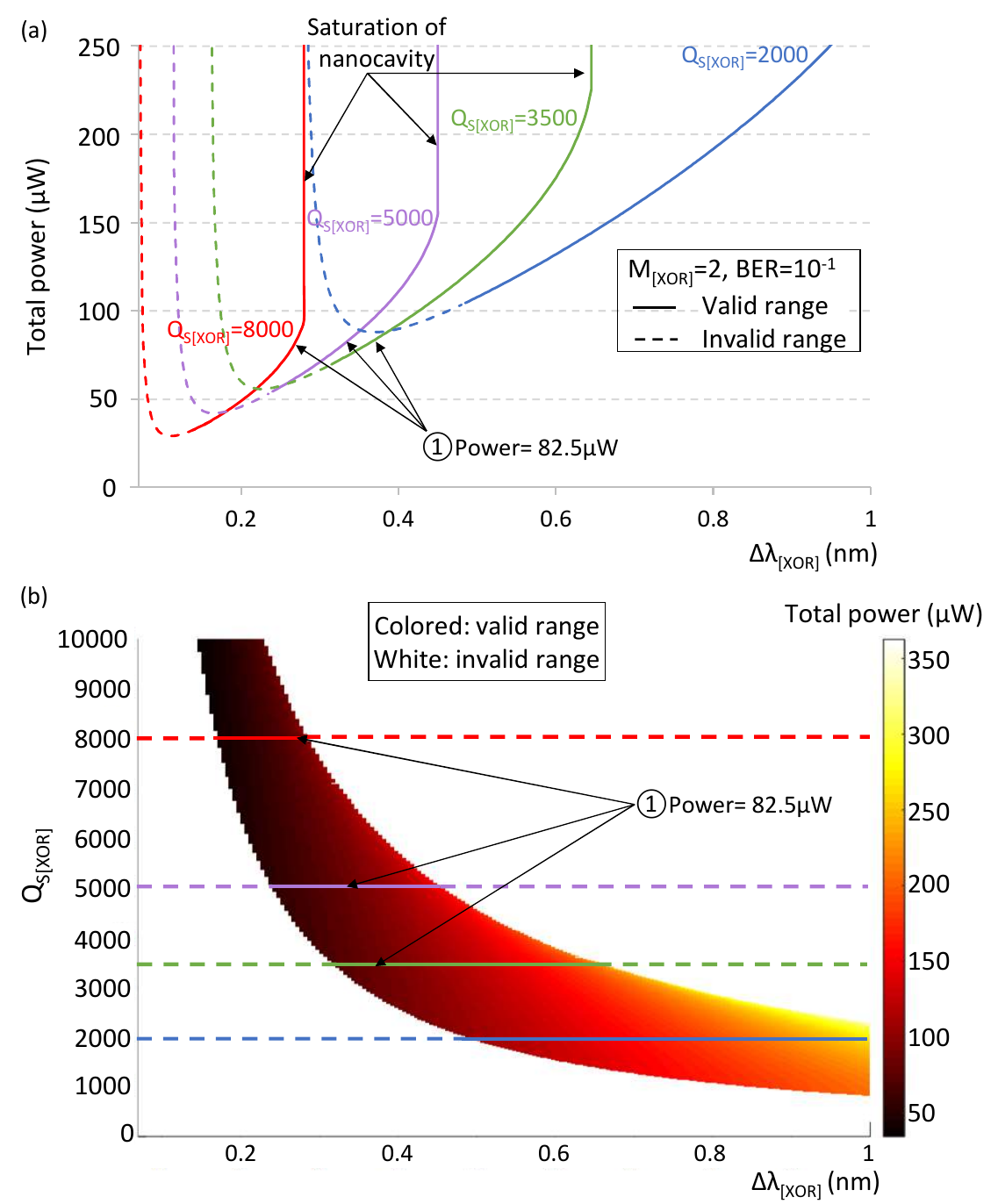}
\caption[Total lasers power of XOR gate]{Total lasers power of XOR gate assuming $BER=10^{-1}$, $M_{[XOR]}=2$ and: (a) $Q_{S[XOR]}=2000$, 3500, 5000, and 8000. (b) $Q_{S[XOR]}$ ranges from 1 to 10000.}
\label{fig:XOR_result}
\end{figure}

As previously defined, an XOR gate is composed of two cascaded nanocavities with the same $Q$ factor but with resonances separated by $\Delta\lambda_{[XOR]}$. We assume $M_{[XOR]}$=2 and $Q_{S[XOR]}$=[2000; 3500; 5000; 8000]. Figure~\ref{fig:XOR_result}(a) illustrates the total power consumption for $\Delta\lambda_{[XOR]}$ ranging from 0 to 1nm and for a targeted $BER=10^{-1}$. As it can be seen in the figure, $Q_{S[XOR]}$=8000 and 2000 lead to a valid $\Delta\lambda_{[XOR]}$ range of [0.17-0.28]nm and [0.45-1.13]nm, respectively, and involve a total power consumption ranging from 39$\mu$W to 94$\mu$W and 104$\mu$W to 276$\mu$W, respectively. Hence, the lower $Q_{S[XOR]}$, the larger the valid range of $\Delta\lambda_{[XOR]}$ and the more increases the power overhead. As also can be observed from the figure, a total power=82.5$\mu$W can be obtained for $Q_{S[XOR]}$=8000, 5000, and 3500 under $\Delta\lambda_{[XOR]}$ =0.27nm, 0.335nm, and 0.365nm, respectively (see $\textcircled{1}$). This demonstrates that the same power efficiency can be obtained for different cavities ($Q_{S[XOR]}$) and wavelength detuning ($\Delta\lambda_{[XOR]}$). 

In the following, we explore $Q_{S[XOR]}$ and $\Delta\lambda_{[XOR]}$ with the aim to find design parameters that minimize the $XOR$ power consumption. The results are reported in Figure~\ref{fig:XOR_result}(b). For the sake of clarity, the design parameters corresponding to cavities detailed in Figure~\ref{fig:XOR_result}(a) are highlighted in Figure~\ref{fig:XOR_result}(b) (mark $\textcircled{1}$). As a first observation, we note that the higher $Q_{S[XOR]}$ and the lower $\Delta\lambda_{[XOR]}$, the lower the power consumption, which is due to the reduced amount of energy needed to shift the cavity. Overall, the cavities laser power consumption ranges from 34.7$\mu$W (at $\Delta\lambda_{[XOR]}$ =0.14nm and $Q_{S[XOR]}$=10000) to 398.2$\mu$W (at $\Delta\lambda_{[XOR]}$=1nm and $Q_{S[XOR]}$=2000). As discussed earlier, we use the same parameters for the cavities located in the XOR stage and the first MUX stage. In the following, we explore the remaining design parameters for MUX stages. 
\vspace{-10pt}
\subsection{Design of MUX}
\label{sec:cascaded_MUX}
In the following, we explore the MUX design parameters. For this purpose, we target a $BER=5\times10^{-1}$ at the photodetector, which corresponds to $BER$ at stage $n$=3 of the MUX ($BER_{[MUX,3]}$), and we explore the design space from the first stage to the last stage, by defining the inter-stage $BER$ to be reached. We use the corresponding parameters ($Q_{S[MUX,n]}$, $WLS_{n}$) from stage $n$ to explore the design space of stage $n$+1.
\begin{itemize}[leftmargin=*]
\item \textbf{Stage \textit{n}=1:} we assume 3$\mu$W input signals powers ($OLP_{Input}$) injected in the $XOR$ gates, we also assume the following ranges for $Q$ factors and $WLS_{1}$: $1<Q_{S[MUX,1]}<10000$ and $0<WLS_{1}<1.2$nm. As shown in Figure~\ref{fig:MUX_result}(a), the exploration results in $BER_{[MUX,1]}$ ranges between $10^{-4}$ and $5\times10^{-1}$. As can be seen, a high $Q_{S[MUX,1]}$ leads to more accurate designs. For example, $Q_{S[MUX,1]}$=10000 and 5000 result in $BER_{[MUX,1]}$= [$10^{-4}$ - $4\times10^{-4}$] and [$4\times10^{-4}$ - $5\times10^{-2}$], respectively. Moreover, the higher $WLS_{1}$, the lower $BER_{[MUX,1]}$, which is due to the reduced crosstalk. We choose $Q_{S[MUX,1]}$=10000 and $WLS_{1}$=0.215nm, which lead to the lowest possible $BER$ for the covered design space ($BER_{[MUX,1]}=10^{-4}$). The corresponding transmission is plotted in the caption of Figure~\ref{fig:MUX_result}(a). The data signals propagate at $\lambda_{S[1]}$=1542nm (i.e., baseline wavelength obtained through experimental results) and $\lambda_{S[2]}$=1541.785nm (i.e., baseline wavelength minus the 0.215nm spacing). The detuning of the cavity to $\lambda_{S[2]}$ is obtained with a 32$\mu$W pump power. The selected signal is transmitted to the MUX output with a power of 1.2$\mu$W.

\item \textbf{Stage \textit{n}=2:} We assume the parameters defined in stage $n$=1 (i.e., $Q_{S[MUX,1]}$=10000 and $WLS_{1}$=0.215nm) and we explore the same ranges of values for $Q_{S[MUX,2]}$ and $WLS_{2}$. Figure~\ref{fig:MUX_result}(b) shows the resulting $BER$ at stage $n$=2 ($BER_{[MUX,2]}$), which is overall higher than $BER_{[MUX,1]}$ due to: i) the higher crosstalk induced by additional input signals to process (2 and 4 input signals at $n$=1 and $n$=2, respectively) and ii) the lower received data signal power (3$\mu$W and 1.2$\mu$W at $n$=1 and $n$=2, respectively). We target $10^{-2}$ for $BER_{[MUX,2]}$, which we obtain with $Q_{S[MUX,2]}$=1900 and $WLS_{2}$=1.19nm (for a 210$\mu$W pump power). The resulting transmission is shown in the caption. In addition to the input signals at $\lambda_{S[1]}$ and $\lambda_{S[2]}$, we inject signals at $\lambda_{S[3]}$=1540.81nm and $\lambda_{S[4]}$=1540.595nm: the distance between $\lambda_{S[3]}$ and $\lambda_{S[4]}$ is 0.215nm and the distance between $\{\lambda_{S[1]}$, $\lambda_{S[2]}\}$ and $\{\lambda_{S[3]}$, $\lambda_{S[4]}\}$ is 1.19nm.

\begin{figure}[h]
\centering
\captionsetup{justification=centering}
\includegraphics[width=\textwidth]{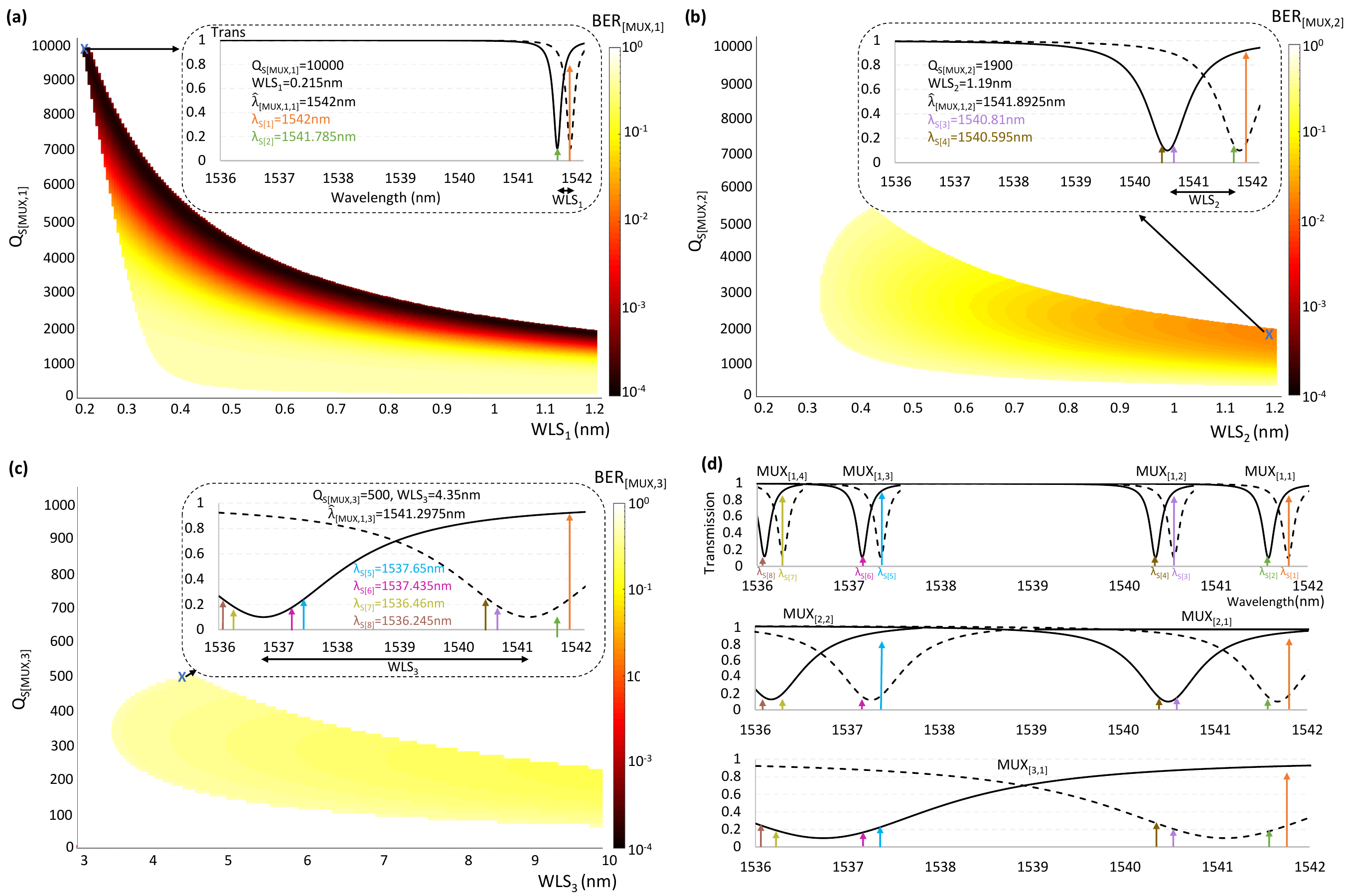}
\caption[Achievable $BER$ at each stage for nanocavities with $M_{[MUX]}=2$]{Achievable $BER$ at each stage for nanocavities with $M_{[MUX]}=2$: (a) Stage $n=1$ with $1 < Q_{S[MUX,1]} < 10000$ and $0< WLS_{1} < 1.2nm$. (b) Stage $n=2$ with $1 < Q_{S[MUX,2]} < 10000$ and $0 < WLS2 < 1.2nm$. (c) Stage $n=3$ with $1 < Q_{S[MUX,3]} <1000$ and $3 < WLS_{3} < 10nm$. (d) The transmission of the MUXs at different stages.}
\label{fig:MUX_result}
\end{figure}

\item \textbf{Stage \textit{n}=3:} The design of the MUX at stage $n$=3 ($MUX_{[1,3]}$) is explored assuming $Q_{S[MUX,2]}$=1900 and $WLS_{2}$=1.19nm. As reported in Figure~\ref{fig:MUX_result}(c), $Q_{S[MUX,3]}$=500 and $WLS_{3}$=4.35nm lead to the targeted $5\times10^{-1}$ $BER$. The 8 signals received by $MUX_{[1,3]}$ and the corresponding cavity transmission are illustrated in the caption. The selected value for $WLS_{3}$ leads to $\lambda_{S[5]}$=1537.65nm, $\lambda_{S[6]}$=1537.435nm, $\lambda_{S[7]}$=1536.46nm, and $\lambda_{S[8]}$=1536.245nm. The selection of signals $\lambda_{S[4-8]}$ is achieved by applying a 670$\mu$W pump power.

\end{itemize}

As it has been observed, the design space considerably shrinks from a stage to another, which is mostly due to the increasing number of signals to process, as shown in Figure~\ref{fig:MUX_result}(d). This calls for increasing wavelength spacing and thus reducing $Q_{S[gate]}$. As a matter of fact, we found that the highest possible $Q$ factor should be preferred for the design of the XOR gates. Regarding the error rate, which inevitably increases as signals propagate through the stages, it can be overcome by increasing the power laser and the $BSL$, as discussed in the sequel. Overall, the optimization of the architecture would benefit from heuristics to explore the design space, which is out of the scope of the work, but we plan to investigate this in our future work.

\subsection{Application-level Design Comparison}
In the following, we evaluate the application level computing accuracy, energy consumption and processing time of the architecture. For a comparison purpose, we assume injected input power signals at 3$\mu$W and 4$\mu$W, and we target $5\times10^{-1}$ and $10^{-1}$ $BER$, respectively. By following the algorithm defined in Section~\ref{nanocavity_design_parameters}, we obtain Design A and Design B, for which the $Q$ factors and wavelength spacings are reported in Table~\ref{table:designA_B}.

\begin{figure}[!t]
\centering
\includegraphics[width = 0.9\textwidth]{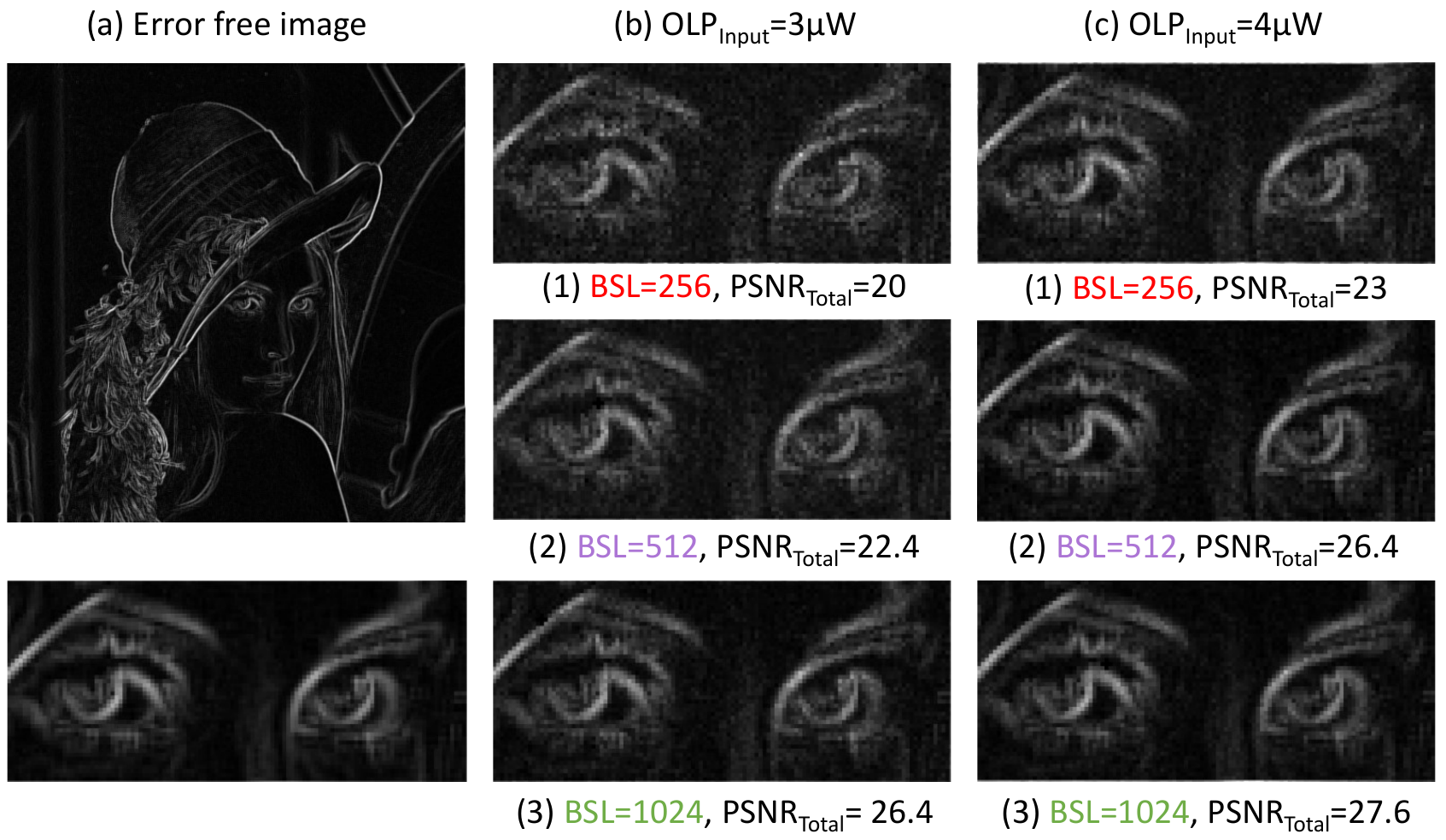}
\captionsetup{justification=centering}
\caption[Processed image: (a) error free, and \textit{PSNR\textsubscript{Total}}]{Processed image: (a) error free, and \textit{PSNR\textsubscript{Total}} for (b) \textit{OLP\textsubscript{Input}}=3µW and (c) \textit{OLP\textsubscript{Input}}=4µW assuming $BSL$=256, 512, and 1024.}
\label{fig:accuracy_image}
\end{figure}

\begin{table}[!t]
\setlength{\tabcolsep}{1.5pt} 
\normalsize
\centering
\captionsetup{justification=centering}
\caption{Device/system-level parameters, and performance of two designs target \textit{PSNR\textsubscript{Total}}=26.4.}
\label{table:designA_B}
\begin{tabular}{cc|c|c|}
\cline{3-4}
\textbf{}                                                                                                           &                                                                         & \multicolumn{2}{c|}{\textbf{\begin{tabular}[c]{@{}c@{}}Computing accuracy\\ \textit{PSNR\textsubscript{Total}}=26.4\end{tabular}}} \\ \cline{3-4} 
\textbf{}                                                                                                           &                                                                         & \textbf{Design A}                                    & \textbf{Design B}                                  \\ \hline
\multicolumn{1}{|c|}{}                                                                                              & \textit{OLP\textsubscript{Input}}                                                                & 3$\mu$W                                                  & 4$\mu$W                                                \\ \cline{2-4} 
\multicolumn{1}{|c|}{\multirow{-2}{*}{\textbf{\begin{tabular}[c]{@{}c@{}}Input\\ parameters\end{tabular}}}}         & $BSL$                                                                     & {\color[HTML]{009901} 1024}                          & {\color[HTML]{987CF1} 512}                         \\ \hline
\multicolumn{1}{|c|}{}                                                                                              & \textit{Q\textsubscript{S[XOR]}}=\textit{Q\textsubscript{S[MUX,1]}}                                               & 10000                                                & 7700                                               \\ \cline{2-4} 
\multicolumn{1}{|c|}{}                                                                                              & \textit{Q\textsubscript{S[MUX,2]}}                                                           & 1900                                                 & 1600                                               \\ \cline{2-4} 
\multicolumn{1}{|c|}{\multirow{-3}{*}{\textbf{\begin{tabular}[c]{@{}c@{}}Device \\ parameters\end{tabular}}}}       & \textit{Q\textsubscript{S[MUX,3]}}                                                           & 500                                                  & 200                                                \\ \hline
\multicolumn{1}{|c|}{}                                                                                              & \textit{WLS\textsubscript{1}} (nm)                                                               & 0.215                                                & 0.275                                              \\ \cline{2-4} 
\multicolumn{1}{|c|}{}                                                                                              & \textit{WLS\textsubscript{2}} (nm)                                                               & 1.19                                                 & 1.41                                               \\ \cline{2-4} 
\multicolumn{1}{|c|}{\multirow{-3}{*}{\textbf{\begin{tabular}[c]{@{}c@{}}System-level \\ parameters\end{tabular}}}} & \textit{WLS\textsubscript{3}} (nm)                                                               & 4.35                                                 & 11.3                                               \\ \hline
\multicolumn{1}{|c|}{}                                                                                              & $BER$                                                                     & 5$\times10^{-1}$                                               & $10^{-1}$                                               \\ \cline{2-4} 
\multicolumn{1}{|c|}{}                                                                                              & \begin{tabular}[c]{@{}c@{}}Energy consumption\\ (nJ/pixel)\end{tabular} & 9                                                  & 8.5                                               \\ \cline{2-4} 
\multicolumn{1}{|c|}{\multirow{-3}{*}{\textbf{Performance}}}                                                        & \begin{tabular}[c]{@{}c@{}}Processing time\\ (ns/pixel)\end{tabular}    & 1024                                                & 512                                               \\ \hline
\end{tabular}%
\vspace{-10pt}
\end{table}

In order to evaluate the computing accuracy at the application level, we process $512\times512$ pixels images assuming \textit{BSL}=256, 512, and 1024. This results in three designs for each set of parameters, as illustrated in Figure~\ref{fig:accuracy_image}(b) and (c). The error is calculated with respect to the error free image shown in Figure~\ref{fig:accuracy_image}(a). As expected, the accuracy increases with \textit{BSL}. For instance, in Figure~\ref{fig:accuracy_image}(b), \textit{PSNR\textsubscript{Total}} is reduced from 20 to 26.4 when \textit{BSL} is increased from 256 to 1024. Furthermore, the use of \textit{BSL}=1024 for Design A and \textit{BSL}=512 for Design B results in \textit{PSNR\textsubscript{Total}}=26.4, thus leading to opportunities to explore power and processing time tradeoffs. For this purpose, we evaluate the energy per computed pixel assuming 10ps pump pulse width under 1GHz repetition rate and 20\% lasing efficiency. As reported in Table~\ref{table:designA_B}, Design B results in 5.6\% energy saving and $2\times$ reduction in processing time compared to Design A. This indicates that for the assumed set of device parameters, \textit{BSL} has a higher negative impact on energy consumption compared to \textit{BER} due to the higher static energy. Therefore, a small \textit{BSL} is preferred for higher energy efficiency and faster processing architecture. Furthermore, while a higher injected input signal power contributes to reduce the \textit{BER}, it also significantly reduces the design space due to the higher crosstalk. This calls for cavities with a higher figure of merits ($M_{[gate]}$), as discussed in the following.

\section{Discussion and Future Work}

In this work, we have provided a quantitative analysis of our optical stochastic computing based on consolidated photonic technologies. We have introduced a novel photonic device, which allows implementing all-optical NOT gate, XOR gate and MUX. We have also demonstrated the design of all-optical cascaded gates using nanocavities. In order to reduce the energy consumption, nanocavities of higher \textit{Q} factors are required. As seen from the results, the use of high \textit{Q} nanocavities implies a limited wavelength detuning, which drastically reduce the design space. This calls for nanocavities with high figure of merits (i.e., $M_{[gate]}>2$), which we plan to fabricate to demonstrate their full potential in large scale designs. Indeed, while $M_{[gate]}>1$ is generally observed in PhC resonators as the modes are differently spatially confined, larger $M_{[gate]}$ should rely on a tailor-made conception of the PhCs structures allowing two modes with a large difference of \textit{Q} factors. This will be achieved, for instance, by optimizing the coupling strength of each of these modes to the waveguide. In order to accurately control the operating wavelength of each gate, micro-heaters~\cite{padmaraju2013integrated} will be implemented on the nanocavities, which will be taken into account in our energy model. At system level, the exploration of figure of merits will considerably increase the design space to explore, which calls for heuristic algorithms we will develop to efficiently optimize architectures.

Our future work aims to design fully optical stochastic computing architectures. This will allow conducting a comprehensive comparison with binary conventional and stochastic computing CMOS-based architectures~\cite{ranjbar2015using, joe2019efficient}. For this purpose, the design of all-optical SNG will be investigated. An all-optical implementation of the SNG would greatly improve the perspectives of our approach since they would allow to avoid the use of LFSR, comparators and modulators. An all-optical SNG could be based on the chaotic dynamics of semiconductor diode laser~\cite{uchida2008fast}. The statistical qualities of these sources have been validated against a variety of statistical tests such as NIST~\cite{reidler2009ultrahigh}. They generated random streams of data at a rate of 10 GHz or more~\cite{sciamanna2015physics}. The energy consumed by chaotic lasers could be reduced by replacing them with integrated lasers~\cite{wan20171} or recently demonstrated nanolasers~\cite{crosnier2017hybrid}. Furthermore, they are readily integrated on a silicon photonic chip, similar to the optical gates described here, with typical electric power threshold for lasing below 1mW. All-optical stochastic computing architectures will be used to design complex accelerators. We will first design architectures with larger filter patterns and we will then address the design of FIR and IIR filters. Eventually, we plan to develop a tool allowing to synthesize and optimize optical stochastic accelerators from high level descriptions of combinational applications.

\section{Conclusion}
In this work, we investigated the use of PhC nanocavity to design a stochastic computing architecture. We proposed a generic transmission model for the nanocavity, which showed a good correlation with experimental measurements for a NOT gate of $Q_{P[NOT]}$=2400 and $M_{[NOT]}$=0.5, hence validating the proposed model. The results showed that we can reach an \textit{ER}= 6.9dB for $\delta \lambda_{[NOT]}$=0.35nm when 34.7$\mu$W power is injected. We used the model to design XOR gate and MUX of different device parameters. We showed that an XOR gate of $Q_{S[XOR]}$=10000 and wavelength detuning equals to 0.14nm leads to 34.7$\mu$W power consumption. We designed an edge detection filter that relies on the proposed nanocavities-based XOR gate and MUX. We showed that it is possible to implement the filter using a design of \textit{Q} factors=7700 for XOR gates and 7700, 1600, and 200 for MUXs. At the application-level, images were processed for various lasers power and \textit{BSL}. The results showed that the assumed set of device parameters, \textit{BSL} has a higher negative effect on the energy consumption compared to \textit{BER}. The resulting architecture showed 8.5nJ/pixel energy consumption and 512ns/pixel processing time. All these observations raised the need to fabricate nanocavities with a higher figure of merits ($M_{[gate]}$) to increase the design space of the gates, which we plan to investigate in our future work. Other perspectives include the design of optical SNG and de-randomizer circuits, which will allow us to show the potential of integrated optics in accelerating stochastic computing architectures.

\bibliographystyle{unsrt}

\bibliography{mabiblio}
\end{document}